\documentclass[showkeys,nofootinbib,longbibliography,11pt]{revtex4-2}

\usepackage[pdftex]{graphicx}
\usepackage{hyperref}

\usepackage{amsmath,amssymb}
\usepackage{bm} 
\usepackage{mathtools}
\usepackage[caption=false]{subfig}

\usepackage{color}
\usepackage{ulem}

\begin{document}

\title{SWKB Quantization Condition for Conditionally Exactly Solvable
Systems and the Residual Corrections}
\author{Yuta Nasuda}
\email{y.nasuda.phys@gmail.com}
\affiliation{Department of Physics, Graduate School of Science and Technology, Tokyo University of Science, Noda, Chiba 278-8510, Japan}

\author{Nobuyuki Sawado}
\email{sawadoph@rs.tus.ac.jp}
\affiliation{Department of Physics, Graduate School of Science and Technology, Tokyo University of Science, Noda, Chiba 278-8510, Japan}

\date{\today}

\begin{abstract}
The SWKB quantization condition is an exact quantization condition for the conventional shape-invariant potentials.
On the other hand, this condition equation does not hold for other known solvable systems. 
The origin of the (non-)exactness is understood in the context of the quantum Hamilton--Jacobi formalism.
First we confirm the statement and show inexplicit properties numerically for the case of the conditionally exactly solvable systems by Junker and Roy.
The SWKB condition clearly breaks for this case, but the condition equation is restored within a certain degree of accuracy.
We propose a novel approach to evaluate the residual by perturbation, intending to explore what the correction terms for the SWKB condition equation look like.
\end{abstract}


\maketitle

\section{Introduction}
\label{sec:Intro}

The supersymmetric quantum mechanics (SUSY QM)~\cite{Witten:1981nf,witten1982constraints,Cooper:1994eh} is a powerful tool for analyzing spectra of many potential problems in quantum mechanics with one space coordinate. 
The study usually deals with special classes of potentials, \textit{i.e.}, the solvable potentials, which are often models for the realistic bound-state problems (\textit{e.g.}, Ref.~\cite{infeld1951factorization}).
The well-known examples are systems possessing the shape invariance (SI)~\cite{gendenshtein1983derivation}.
It is a sufficient (but not a necessary) condition for the exact solvability of the Schr\"{o}dinger equation.
There are numerous studies concerning the shape-invariant (SI) potentials; 
the authors compared some solvable potentials with the related SI ones~\cite{cooper1987relationship,barclay1993new} 
to inquire the origin of the exactness.  
Also, several novel types of SI potentials~\cite{barclay1993new,Quesne_2008,Odake:2009zv,Odake:2010zz,Sasaki_2010,Odake:2011jj,bagchi2005deformed}
have been proposed so far.

In 1980's, Comtet and co-authors proposed a WKB-like integral form of  quantization condition in the context of SUSY QM:
\begin{equation}
I_{\mathrm{SWKB}}=n\pi\hbar ~,~~~
I_{\mathrm{SWKB}} \coloneqq
\int_{a_{\mathrm{L}}}^{a_{\mathrm{R}}} \sqrt{\mathcal{E}_n - W(x)^2}\,\mathrm{d}x ~,~~~
n \in \mathbb{Z}_{\geqslant 0} ~,
\label{eq:SWKBqc1}
\end{equation}
which is called the supersymmetric WKB (SWKB) quantization condition~\cite{Comtet:1985rb}.
Here, $W(x)$ is the superpotential. 
This condition seems to have a deep physical insight, for it successfully reproduces the exact bound-state spectra for all conventional SI potentials as was demonstrated by Dutt \textit{et al.}~\cite{Dutt:1986pi}. 
Many researchers at the period thought that the SWKB condition became exact if and only if the system possesses SI.

After that, several efforts have been made for this conjecture.
Khare and Varshni demonstrated the non-exactness of the SWKB condition for the Ginocchio potential~\cite{ginocchio1984class} 
and a potential iso-spectral to the one-dimensional harmonic oscillator (1-d H.O.)~\cite{abraham1980changes}, both of which are exactly solvable but are not SI~
\cite{Khare:1989zy}. 
DeLaney and Nieto showed that the Abraham--Moses systems~\cite{abraham1980changes}, 
which are another class of solvable potentials without SI, do not satisfy the SWKB condition~\cite{delaney1990susy}.
In our previous letter~\cite{nasuda2021numerical}, we have provided yet another example; 
the Krein--Adler systems~\cite{Kre57,adler1994modification}, which are solvable but not SI, do not satisfy the SWKB condition.
Obviously, the exactly solvable systems without SI always break the exactness of SWKB condition.
Moreover, it has been confirmed that a novel class of potentials with SI, called the multi-indexed systems~\cite{Quesne_2008,Odake:2009zv,Odake:2010zz,Sasaki_2010,Odake:2011jj},  does not satisfy the SWKB condition~\cite{Bougie:2018lvd,nasuda2021numerical}.
This clearly indicates that the SI of a system is not a sufficient condition for the exactness of the SWKB condition.
Now, it is concluded that SI does not account for the exactness of the SWKB condition.

In 1996, Bhalla \textit{et al.} gave a qualitative explanation for the (non-)exactness of the SWKB condition by means of the quantum Hamilton--Jacobi (QHJ) formalism~\cite{bhalla1996exactness,bhalla1997quantum}.
The QHJ formalism was first developed by Leacock and Padgett~\cite{leacock1983hamilton_action,leacock1983hamilton}, and is another approach toward potential problems.
By employing this, one can obtain the exact bound-state spectra without any information of the eigenstate wave functions.
The energy eigenvalues are given exactly by the quantization of a quantum analogue of the action variable, 
which is similar to the relation between the frequency and the action variable in classical mechanics~\cite{goldstein2002classical}.
The quantization condition for the quantum action variable is
\begin{equation}
J_{\mathrm{QHJ}} = n\hbar ~,~~~
J_{\mathrm{QHJ}} \coloneqq \frac{1}{2\pi} \oint_C p(x;\mathcal{E}) \,\mathrm{d}x ~,~~~
n \in \mathbb{Z}_{\geqslant 0} ~, 
\label{eq:QHJqc1}
\end{equation}
where $p(x;\mathcal{E})$ is called the quantum momentum function (QMF).
One can construct the QMF from the singularity structure and the QHJ equation, and then obtain the wave function explicitly~\cite{ranjani2004bound}.

Bhalla \textit{et al.} pointed out that Eq.~\eqref{eq:QHJqc1} is equivalent to the SWKB quantization condition \eqref{eq:SWKBqc1} for the conventional SI systems.
They concluded that the SWKB condition becomes an exact quantization condition 
when the integrand (in which the real variable $x$ is extended to a complex variable $x$) has the common singularity nature with the QMF~\cite{bhalla1996exactness,bhalla1997quantum}.  
This indicates that for the case of the conventional SI potentials, the complex integrations for the QMFs can be deformed into simple WKB-like integrations along the real axis; while for the other potentials, the deformation is not so straightforward, and one has to carry out complicated counter integrals.
The SWKB condition, on the other hand, gives approximate results by a simple integration along with the real axis, where the SWKB scheme takes some advantages.

Incompatibility of these singularities thus concern the higher order corrections of the SWKB condition, \textit{i.e.}, the residual $\Delta \coloneqq \pi J_{\mathrm{QHJ}} - I_{\mathrm{SWKB}}$.
In Ref.~\cite{bhalla1997quantum}, $\Delta$ is estimated with the residues of the poles of the SWKB integrand that essentially is different from those of the QMF.
From this point of view, they got a quite natural explanation about why the SWKB condition of some potentials cannot reproduce the exact bound-state spectra.
However, yet another difficulty exists in due course; one cannot always calculate the poles analytically in some potentials and the integration is not straightforward in these cases.
For the method of evaluating $\Delta$, several approaches have already formulated so far~\cite{adhikari1988higher,dutt1991supersymmetry,ranjani2012exceptional,Bougie:2018lvd}.

In order to achieve a deeper understanding of the SWKB condition, we study $\Delta$ somewhat different point of view.
For a given potential, suppose we have a quantized integral and  
the SWKB integral $I_{\mathrm{SWKB}}$ is a leading order of an expansion with a parameter which is not yet fixed (the ``unknown parameter''). 
All the higher order corrections are gathered in $\Delta$ and then,  an ``exact'' SWKB condition is formally expressed as follows: 
$I_{\mathrm{SWKB}} + \Delta = n\pi\hbar$.
Note that for the conventional SI systems, $\Delta = 0$.
The evaluation of $\Delta$ is important because it brings us a good intuition for the meaning of the condition, and also one could obtain an ``exact'' SWKB formula for a given potential, which is certainly a final goal of the study on the SWKB condition.

Now, it is almost certain that the breaking of the SWKB condition is caused by properties of the system in question, which is to be described by some model parameters.
The question here is how the value of $I_{\mathrm{SWKB}}$, or $\Delta$, changes as the model parameters vary.
In our previous letter~\cite{nasuda2021numerical}, we have conjectured that, in terms of the energy eigenvalues, the whole distribution of them guarantees the exactness of the SWKB condition.
This conjecture would be a good starting point of our study on $\Delta$.
Here, we analyze a system that is related to the conventional SI systems by continuous parameters describing the modification of energy spectrum.
In our analysis, we employ such a system known as the conditionally exactly solvable (CES) system by Junker and Roy~\cite{junker1998conditionally}.
This class of potentials have two additional model parameters;
one describes a constant shift of excited energies while the other is responsible for an iso-spectral deformation.  
This allows us to carry out our analysis by a perturbative approach.

As was mentioned above, the investigation of $\Delta$ is a significant issue and has several future applications.
We add two more here.
First, the SWKB condition can give an approximation formula for the energy in the same manner as the WKB formalism (See Appendix~\ref{sec:Appx}).
Case-by-case analysis tells that the SWKB formalism gives better estimations for the energy eigenvalues than the WKB for many cases.
The analysis on $\Delta$ will enable us to obtain an even more accurate formula.
Second, although the conjecture that the SWKB condition is an exact condition if and only if we deal with the conventional SI potentials is solved negatively, there is still a chance the condition is somehow involved with the solvability of the Schr\"{o}dinger equation.
By getting to know how $\Delta$ behaves around $\Delta=0$, the role that the SWKB condition plays in the solvability of the Schr\"{o}dinger equation could be unveiled.

In the next section, we give a brief account on the two quantization conditions and the CES systems by Junker and Roy.
In the first half of Sec.~\ref{sec:SWKB}, we compute both integrals in Eqs.~\eqref{eq:SWKBqc1} and \eqref{eq:QHJqc1} for the CES system numerically to reveal inexplicit properties of the system and identify the origin of the non-exactness of the SWKB condition.
At this point, one realizes the breaking behavior of the SWKB condition can be treated by perturbative approach. 
The latter half of the section is devoted for a special case of the CES system, where one can perform the computations analytically in part. 
Sec.~\ref{sec:Delta} investigate the non-exactness of the SWKB condition by perturbation based on a series expansion of the SWKB integrand.
Then, in the last section, we conclude and state some future directions.

\section{Preliminaries}
\label{sec:Pre}

\subsection{SWKB quantization condition}

In the standard WKB formalism, the quantization condition is given in terms of  the potential $V(x)$, which is formally given by using the ground-state wave function $\psi_0(x)$;
\begin{equation}
V(x) = \hbar^2 \left[ \left( \frac{\mathrm{d}}{\mathrm{d}x}\ln\psi_0(x) \right)^2 + \frac{\mathrm{d}^2}{\mathrm{d}x^2}\ln\psi_0(x) \right]
\equiv W(x)^2 - \hbar W'(x) ~,
\label{eq:V}
\end{equation}
where the  superpotential is $W(x) \coloneqq -\hbar\frac{\mathrm{d}}{\mathrm{d}x}\ln\psi_0(x)$, and $'$ denotes the first derivative in $x$: $' \equiv \frac{\mathrm{d}}{\mathrm{d}x}$.
We set $2m=1$ but retain $\hbar$ for discussing explicit $\hbar$-dependency throughout this paper.
We note that Eq.~\eqref{eq:V} corresponds to the vanishing ground-state energy: $\mathcal{E}_0 = 0$.

On the other hand, the SWKB quantization condition only concerns the first term in Eq.~\eqref{eq:V}.
The condition equation given by Comtet \textit{et al.}~\cite{Comtet:1985rb} reads
\begin{equation}
I_{\mathrm{SWKB}} = n\pi\hbar ~,~~~
I_{\mathrm{SWKB}} \coloneqq
\int_{a_{\mathrm{L}}}^{a_{\mathrm{R}}} \sqrt{\mathcal{E}_n - W(x)^2}\,\mathrm{d} x ~,~~~
n \in \mathbb{Z}_{\geqslant 0} ~.
\label{eq:SWKB}
\end{equation}
Here, $a_{\mathrm{L}}$, $a_{\mathrm{R}}$ ($a_{\mathrm{L}} < a_{\mathrm{R}}$) are the ``turning points''; $a_{\mathrm{L}}$ and $a_{\mathrm{R}}$ are the two roots of the equation $W(x)^2 = \mathcal{E}_n$.
We call the integral $I_{\mathrm{SWKB}}$ the SWKB integral.
The condition is always exact for the ground state by construction.
Also, it is well-known that the condition is conserved for a class of potentials called the conventional SI potentials~\cite{Dutt:1986pi,gangopadhyaya2020exactness,gangopadhyaya2021exactness}, while it is not for other potentials constructed so far.
An essential aspect with the condition equation is that $\hbar$ can always be factored out from the SWKB integral and the condition is independent of $\hbar$~\cite{nasuda2021numerical}, which was often missed in the literatures.

In some cases, the equation $W(x)^2 = \mathcal{E}_n$ possesses more than two roots:
 $\{ (a_{\mathrm{L},i},a_{\mathrm{R},i});~ i=1,\ldots,N \}$. 
One can employ a prescription of estimating $I_{\mathrm{SWKB}}$ by summing up the SWKB integrals of all $i$~\cite{nasuda2021numerical}:
\begin{equation}
I_{\mathrm{SWKB}} =
\sum_i \int_{a_{\mathrm{L},i}}^{a_{\mathrm{R},i}} \sqrt{\mathcal{E}_n - W(x)^2}\,\mathrm{d} x ~.
\end{equation}
In this paper, we only consider the cases where the equation has at most two solutions for the rigorous study on the non-exactness of the SWKB condition.

\subsection{QHJ formalism and exact quantization condition}
In the QHJ formalism~\cite{leacock1983hamilton_action,leacock1983hamilton}, the quantum momentum function (QMF):
\begin{equation}
p(x;\mathcal{E}) \coloneqq -\mathrm{i}\hbar\frac{\psi'(x)}{\psi(x)} ~,
\label{eq:QMF}
\end{equation}
plays the central role.
The QMF satisfies the following Riccati-type equation called the QHJ equation:
\begin{equation}
p(x;\mathcal{E})^2 - \mathrm{i}\hbar p'(x;\mathcal{E}) = \mathcal{E} - V(x) ~.
\label{eq:QHJeq}
\end{equation}
The quantum action variable $J_{\mathrm{QHJ}}$ is defined as an analogy with the classical one and the quantization condition for the quantum action variable is written as
\begin{equation}
J_{\mathrm{QHJ}} = n\hbar ~,~~~
J_{\mathrm{QHJ}} \coloneqq \frac{1}{2\pi} \oint_C p(x;\mathcal{E}_n) \,\mathrm{d}x ~,~~~
n \in \mathbb{Z}_{\geqslant 0} ~,
\label{eq:QHJ}
\end{equation}
where $C$ is a counterclockwise contour in the complex $x$-plane enclosing the classical turning points $x_{\mathrm{L}},x_{\mathrm{R}}$: $V(x_{\mathrm{L}}) = V(x_{\mathrm{R}}) = \mathcal{E}_n$.
This condition equation is guaranteed by the fact that an $n$-th eigenfunction has $n$ nodes between the two classical turning points (this is known as the oscillation theorem~\cite{landau1981quantum}), which produce the poles of $p(x;\mathcal{E})$ having residue $-\mathrm{i}\hbar$ along with the real axis.
We note that Gozzi discovered the same equation independently of the QHJ formalism~\cite{gozzi1986nodal}.

Suppose the potential $V(x)$ is in the form of Eq.~\eqref{eq:V} and there exist $n$ poles between the classical turning points along with the real axis, one can write the QMF in the following form:
\begin{equation}
p(x;\mathcal{E}_n) = -\mathrm{i}\hbar \left( \frac{\psi_0'(x)}{\psi_0(x)} + \frac{\mathcal{P}_n'(x)}{\mathcal{P}_n(x)} \right)
= -\mathrm{i}\left( -W(x) + \hbar\frac{\mathcal{P}_n'(x)}{\mathcal{P}_n(x)} \right) ~,
\end{equation}
where $\mathcal{P}_n(x)$ is a function such that $\mathcal{P}_n'(x)/\mathcal{P}_n(x)$ has the $n$ poles of residue $1$ at the same point as the nodes of $\psi_n(x)$ along with the real axis, but $\mathcal{P}_n'(x)/\mathcal{P}_n(x)$ can have other poles off the real axis.
Then, the QHJ equation \eqref{eq:QHJeq} reduces to
\begin{equation}
\hbar^2\mathcal{P}_n''(x) - 2\hbar W(x)\mathcal{P}_n'(x) + \mathcal{E}_n\mathcal{P}_n(x) = 0 ~.
\label{eq:Pn_eq}
\end{equation}
Note that $\hbar$-dependency can be removed from this equation after changing variable.
A function $\mathcal{P}_n(x)$ turns out to be the wave function $\psi_n(x)$ divided by the ground-state wave function $\psi_0(x)$.
Here, we perform the similarity transformation of a Hamiltonian using the ground-state wave function:
\begin{equation}
\mathcal{H} = -\hbar^2\frac{\mathrm{d}^2}{\mathrm{d}x^2} + V(x) \to
\widetilde{\mathcal{H}} \coloneqq \psi_0(x)^{-1}\circ\mathcal{H}\circ\psi_0(x) = -\hbar^2\frac{\mathrm{d}^2}{\mathrm{d}x^2} + 2\hbar W(x)\frac{\mathrm{d}}{\mathrm{d}x} ~.
\end{equation}
Assuming that the wave function is of the form $\psi_n(x) \equiv \psi_0(x)\mathcal{P}_n(x)$ with $\mathcal{P}_n(x)$ being some function, the eigenvalue equation for $\mathcal{H}$ becomes that for $\widetilde{\mathcal{H}}$: 
\begin{equation}
\widetilde{\mathcal{H}}\mathcal{P}_n(x) = \mathcal{E}_n\mathcal{P}_n(x) ~,
\end{equation}
which is equivalent to Eq.~\eqref{eq:Pn_eq}.
Note that $\mathcal{P}_n(x)$ is shown to be the classical orthogonal polynomials for the 1-d H.O., the radial oscillator and the P\"{o}schl--Teller potential. 

Considering the QMF of the form Eq.~\eqref{eq:Pn_eq}, the quantum action variable~\eqref{eq:QHJ} becomes 
\begin{equation}
\frac{1}{2\pi}\oint_C p(x;\mathcal{E}_n)\,\mathrm{d}x 
= \frac{1}{2\pi}\oint_C \sqrt{\mathcal{E}_n - W(x)^2 - \hbar^2\left( \frac{\mathcal{P}_n'(x)}{\mathcal{P}_n(x)} \right)' }\,\mathrm{d}x ~.
\end{equation}
As Bhalla \textit{et al.} pointed out, the SWKB condition is exact when the pole structure of the QMF and that of the SWKB integrand coincide outside the contours~\cite{bhalla1996exactness,bhalla1997quantum}.
Note, however, that in order to compare the two quantization conditions the SWKB integral is extended to a complex contour integral:
\begin{equation}
I_{\mathrm{SWKB}} \to
J_{\mathrm{SWKB}} = \frac{1}{2\pi} \oint_{C'} \sqrt{\mathcal{E}_n - W(x)^2}\,\mathrm{d}x ~,
\end{equation}
where $C'$ is a counterclockwise contour enclosing the branch cut of $\sqrt{\mathcal{E}_n - W(x)^2}$ from $a_{\mathrm{L}}$ to $a_{\mathrm{R}}$.
Apparently, $J_{\mathrm{SWKB}}$ equals $J_{\mathrm{QHJ}}$, when $\bigl( \mathcal{P}_n'(x)/\mathcal{P}_n(x) \bigr)'$ does not have any singularity outside the contour $C$, which is to be realized for all conventional SI systems.
Contrary, for other classes of exactly solvable system, where the quantization of $I_{\mathrm{SWKB}}$ is not exact, it is easy to guess $\bigl( \mathcal{P}_n'(x)/\mathcal{P}_n(x) \bigr)'$ has singularities outside the contour, \textit{i.e.}, the pole structure of the QMF and that of the SWKB integrand are different.

\subsection{Conditionally exactly solvable systems}

In this subsection, we follow the CES systems by Junker and Roy, which is based on SUSY QM.
The name ``conditionally exactly solvable'' reflects that the eigenvalues and eigenfunctions are obtained explicitly for some specific choices of potential parameters~\cite{de1993conditionally,dutt1995new}.

We start with the following three conventional SI potentials: the 1-d H.O. (H), the radial oscillator (L), the P\"{o}schl--Teller potential (J).
The potentials are
\begin{equation}
V^{(\ast)}(x) = \left\{
	\begin{array}{llr}
	\omega^2x^2 - \hbar\omega ~, & x \in (-\infty,\infty) ~ & \mathrm{\ast=H} ~, \\[3pt]
	\omega^2x^2 + \dfrac{\hbar^2g(g-1)}{x^2} - \hbar\omega(2g+1) ~, & x \in (0,\infty) ~ & \mathrm{\ast=L} ~, \\[7.5pt]
	\dfrac{\hbar^2 g(g-1)}{\sin^2x} + \dfrac{\hbar^2 h(h-1)}{\cos^2x} - \hbar^2(g+h)^2 ~,~~ & x \in \left( 0,\dfrac{\pi}{2} \right) & \mathrm{\ast=J} ~.
	\end{array}
\right.
\label{eq:cSI_pot}
\end{equation}
The eigenvalues and the corresponding eigenfunctions are
\begin{align}
\mathcal{E}_n^{(\ast)} &= \left\{
        \begin{array}{ll}
        2n\hbar\omega & \mathrm{\ast=H} ~, \\[5pt]
        4n\hbar\omega & \mathrm{\ast=L} ~, \\[5pt]
        4\hbar^2 n(n+g+h) \qquad & \mathrm{\ast=J} ~,
        \end{array}
\right. 
\label{eq:cSI_Ene} \\
\phi_n^{(\ast)}(x) &= \left\{
        \begin{array}{ll}
        \mathrm{e}^{-\frac{\xi^2}{2}} H_n(\xi) & \mathrm{\ast=H} ~, \\[5pt]
        \mathrm{e}^{-\frac{z}{2}}z^{\frac{g}{2}} L_n^{(g-\frac{1}{2})}(z) & \mathrm{\ast=L} ~, \\[5pt]
        (1-y)^{\frac{g}{2}}(1+y)^{\frac{h}{2}} P_n^{(g-\frac{1}{2},h-\frac{1}{2})}(y) \qquad & \mathrm{\ast=J} ~.
        \end{array}
\right.
\end{align}
Here, $H_n$, $L_n^{(\alpha)}$, $P_n^{(\alpha,\beta)}$ are Hermite, Laguerre, Jacobi polynomials respectively, and  $\xi\equiv\sqrt{\omega/\hbar\,}\,x$, $z\equiv \xi^2$ and $y\equiv \cos 2x$.
Note that the SI transformations are (H) $\emptyset$, (L) $g\to g+1$ and (J) $g\to g+1$, $h\to h+1$.

The CES potentials are defined as
\begin{equation}
V^{(\mathrm{C},\ast)}(x) \coloneqq \hbar^2\left[ \left( \frac{\mathrm{d}}{\mathrm{d}x}\ln\left| \frac{\phi_0^{(\ast)}(x)}{u^{(\ast)}(x)} \right| \right)^2 + \frac{\mathrm{d}^2}{\mathrm{d}x^2}\ln\left| \frac{\phi_0^{(\ast)}(x)}{u^{(\ast)}(x)} \right| \right] ~,~~~
\ast = \mathrm{H,L,J} ~,
\end{equation}
where $u^{(\ast)}(x)$ satisfies
\begin{equation}
\hbar^2u^{(\ast)''}(x) + 2\hbar^2 \left( \frac{\mathrm{d}}{\mathrm{d}x}\ln\phi_0^{(\ast)}(x) \right) u^{(\ast)'}(x) - \tilde{b}u^{(\ast)}(x) = 0 ~.
\label{eq:u}
\end{equation}
The eigenvalues and the corresponding eigenfunctions are
\begin{gather}
\mathcal{E}_0^{(\mathrm{C},\ast)} = 0 ~,~~~
\mathcal{E}_n^{(\mathrm{C},\ast)} = \mathcal{E}_n^{(\ast)} + \tilde{b} \quad\text{for $n\geqslant 1$} ~, \\
\psi_0^{(\mathrm{C},\ast)}(x) = \frac{\phi_0^{(\mathrm{\ast})}(x)}{u^{(\ast)}(x)} ~,~~~
\psi_n^{(\mathrm{C,\ast})}(x) = \hbar\left( -\frac{\mathrm{d}}{\mathrm{d}x} - \frac{\mathrm{d}}{\mathrm{d}x}\ln\left| \psi_0^{(\mathrm{C},\ast)}(x) \right| \right)\psi_{n-1}^{(\ast,+)}(x) \quad\text{for $n\geqslant 1$} ~.
\end{gather}
Here, $\psi_n^{(\ast,+)}(x)$ denotes the $n$-th eigenstate wave function of the SUSY-partner system:
\[
\mathcal{H}^{(+)} = -\frac{\mathrm{d}^2}{\mathrm{d}x^2} + V^{(\ast)}(x) + \tilde{b} ~.
\]
Note that from the positivity of the Hamiltonian, $\mathcal{E}_1^{(\mathrm{C,\ast})} > \mathcal{E}_0^{(\mathrm{C,\ast})}=0$, which gives a condition on the parameter $\tilde{b}$.

For the case of $\ast=\mathrm{H}$, 
\begin{equation}
\tilde{b} \equiv b\hbar\omega > -2\hbar\omega ~,
\label{eq:cond1_CESH}
\end{equation} 
and the general solution for Eq.~\eqref{eq:u} is
\begin{equation}
u^{(\mathrm{H})}(x) = \alpha\,{}_1F_1\left( -\frac{b}{4},\frac{1}{2};-\xi^2 \right) + \beta \xi\,{}_1F_1\left( \frac{1}{2}-\frac{b}{4},\frac{3}{2};-\xi^2 \right) ~.
\end{equation}
The parameters $\alpha$ and $\beta$ satisfy the following condition so that the resulting system does not have singularities in the domain:
\begin{equation}
\left|\frac{\beta}{\alpha}\right| < \frac{2\varGamma\left(\frac{b}{4}+1\right)}{\varGamma\left(\frac{b}{4}+\frac{1}{2}\right)} ~.
\label{eq:cond2_CESH}
\end{equation}
For $\ast=\mathrm{L}$, 
\begin{equation}
u^{(\mathrm{L})}(x) = \alpha\,{}_1F_1\left( -\frac{b}{4},\frac{1}{2}-g;-z \right) + \beta z^{g+\frac{1}{2}} {}_1F_1\left( \frac{1}{2}+g-\frac{b}{4},\frac{3}{2}+g;-z \right) ~,
\end{equation}
and the parameters satisfy
\begin{equation}
\tilde{b} \equiv b\hbar\omega > -4\hbar\omega ~,~~~
\alpha > 0 ~,~~~
\frac{\beta}{\alpha} > -\frac{\varGamma\left(\frac{1}{2}-g\right)}{\varGamma\left(\frac{1}{2}-g+\frac{b}{4}\right)}\cdot\frac{\varGamma\left(\frac{b}{4}+1\right)}{\varGamma\left(\frac{3}{2}+g\right)} ~.
\end{equation}
For $\ast=\mathrm{J}$,
\begin{multline}
u^{(\mathrm{J})}(x) = \alpha\,{}_2F_1\left( -\frac{g}{2}-\frac{h}{2}-\frac{\sqrt{(g+h)^2-b}}{2},-\frac{g}{2}-\frac{h}{2}+\frac{\sqrt{(g+h)^2-b}}{2},\frac{1}{2}-h;\frac{1+y}{2} \right) \\
\qquad + \beta y^{2h+1}{}_2F_1\left( \frac{1}{2}-\frac{g}{2}+\frac{h}{2}-\frac{\sqrt{(g+h)^2-b}}{2},\frac{1}{2}-\frac{g}{2}+\frac{h}{2}+\frac{\sqrt{(g+h)^2-b}}{2},\frac{3}{2}+h;\frac{1+y}{2} \right) ~,
\end{multline}
and
\begin{align}
&\tilde{b} \equiv b\hbar^2 > -4\hbar^2(g+h+1) ~, \nonumber \\
&\alpha-\beta > 0 ~, ~~~
\frac{\beta}{\alpha} > - \frac{\varGamma\left(
\frac{1}{2}-h \right)}{\varGamma\left(
\frac{3}{2}+h \right)} \cdot \frac{\varGamma\left(1+\frac{g}{2}+\frac{h}{2}+\frac{\sqrt{(g+h)^2-b}}{2}\right)\varGamma\left(1+\frac{g}{2}+\frac{h}{2}-\frac{\sqrt{(g+h)^2-b}}{2}\right)}{\varGamma\left(\frac{1}{2}+\frac{g}{2}-\frac{h}{2}+\frac{\sqrt{(g+h)^2-b}}{2}\right)\varGamma\left(\frac{1}{2}+\frac{g}{2}-\frac{h}{2}-\frac{\sqrt{(g+h)^2-b}}{2}\right)} ~.
\end{align}
Hereafter, we fix $\alpha=1$ without loss of generality.
Thus, as was mentioned above, the CES systems have two model parameters; $b$ is responsible for the energy shift while $\beta$ is a parameter describing an iso-spectral deformation of the system.
The case where $b=\beta=0$ is identical to the conventional SI system.

Junker and Roy pointed out in Ref.~\cite{junker1998conditionally} that the CES systems 
of (H) $b=4N$, $\beta=0$, (L) $b=8N$, $\beta=0$ with $N\in\mathbb{Z}_{>0}$ are obtained by
the Krein--Adler transformation of the corresponding conventional SI potentials.
Note that for the case of $\ast=\mathrm{J}$ there has no such correspondence.

The Krein--Adler transformation concerns the deletion of the eigenstates of the original exactly solvable system whose indices are designated by $\mathcal{D}$.
Generally, one can take $\mathcal{D}=\{ d_1, d_1+1 < d_2, d_2+1 < \cdots < d_M,d_M+1\}$ with $\{ d_i\in\mathbb{Z}_{>0}; i=1,\ldots,M \}$, but we consider the deletion of eigenstates indicated by $2N$ consecutive integers, \textit{i.e.}, $\mathcal{D}=\{d, d+1,\ldots,d+2N-1\}$ in this paper. 
Moreover, we restrict ourselves mainly to $N=1$ for simplicity.
The potential for the Krein--Adler systems is
\begin{equation}
V^{\mathrm{(K,\ast)}}(x) \coloneqq V^{(\ast)}(x) - 2\hbar^2\frac{\mathrm{d}^2}{\mathrm{d}x^2}\ln\left|\mathrm{W}\left[\phi_d^{(\ast)},\phi_{d+1}^{(\ast)}\right](x)\right| ~,~~~
\ast = \mathrm{H, L, J} ~,
\label{eq:KA}
\end{equation}
in which $\mathrm{W}[f_1,\ldots,f_n](x) \equiv \det\left( f_k^{(j-1)}(x) \right)_{1\leqslant j,k \leqslant n}$ is the Wronskian.
The eigenvalues and the corresponding eigenfunctions are
\begin{equation}
\mathcal{E}_{\mathcal{D};n}^{\mathrm{(K,\ast)}} = \mathcal{E}_{\breve{n}}^{\mathrm{(\ast)}} ~,~~~
\psi_{\mathcal{D};n}^{\mathrm{(K,\ast)}}(x) = \frac{\mathrm{W}\left[\phi_d^{(\ast)},\phi_{d+1}^{(\ast)},\phi_{\breve{n}}^{(\ast)}\right](x)}{\mathrm{W}\left[\phi_d^{(\ast)},\phi_{d+1}^{(\ast)}\right](x)} ~.
\label{eq:KAene}
\end{equation}
Here, $\breve{n}$ is defined as
\begin{equation}
\breve{n} \coloneqq \left\{
	\begin{array}{l}
	n \qquad (0 \leqslant n \leqslant d-1) \\
	n+2 \qquad (n \geqslant d)
	\end{array}
\right. 
\end{equation}
with $n$ being the number of nodes.

\section{Non-exactness of SWKB condition}
\label{sec:SWKB}

\subsection{SWKB condition for CES systems}

The SWKB integrals for the CES systems are
\begin{equation}
I_{\mathrm{SWKB}} 
= \int_{a_{\mathrm{L}}}^{a_{\mathrm{R}}} \sqrt{\mathcal{E}_n^{\mathrm{(C,\ast)}} - \left( \hbar\frac{\mathrm{d}}{\mathrm{d}x}\ln\left|\psi_0^{\mathrm{(C,\ast)}}(x)\right| \right)^2}\,\mathrm{d}x ~.
\label{eq:SWKB_CES}
\end{equation}
For the case of $\ast=\mathrm{H}$, Eq.~\eqref{eq:SWKB_CES} reduces to
\begin{equation}
I_{\mathrm{SWKB}} 
= \hbar\int_{a'_{\mathrm{L}}}^{a'_{\mathrm{R}}} \sqrt{2n + b - \widetilde{W}^{(\mathrm{C,H})}(\xi)^2}\,\mathrm{d}\xi
\equiv \hbar I^{(\mathrm{C,H})} ~,~~~
\widetilde{W}^{(\mathrm{C,H})}(\xi) \equiv -\frac{\mathrm{d}}{\mathrm{d}\xi}\ln\left|\psi_0^{\mathrm{(C,H)}}(x)\right| ~,
\label{eq:SWKB_CESH}
\end{equation}
and for the cases of $\ast=\mathrm{L, J}$, 
\begin{equation}
I_{\mathrm{SWKB}} 
= \hbar\int_{a'_{\mathrm{L}}}^{a'_{\mathrm{R}}} \sqrt{n + \frac{b}{4} - \widetilde{W}^{(\mathrm{C,L})}(z)^2}\,\frac{\mathrm{d}z}{\sqrt{z}}
\equiv \hbar I^{(\mathrm{C,L})} ~,~~~
\widetilde{W}^{(\mathrm{C,L})}(z) \equiv -\sqrt{z}\,\frac{\mathrm{d}}{\mathrm{d}z}\ln\left|\psi_0^{\mathrm{(C,L)}}(x)\right| ~,
\label{eq:SWKB_CESL}
\end{equation}
\begin{multline}
I_{\mathrm{SWKB}} 
= \hbar\int_{a'_{\mathrm{L}}}^{a'_{\mathrm{R}}} \sqrt{4n(n+g+h) + b - \widetilde{W}^{(\mathrm{C,J})}(y)^2}\,\frac{\mathrm{d}y}{2\sqrt{1-y^2}}
\equiv \hbar I^{(\mathrm{C,J})} ~, \\
\widetilde{W}^{(\mathrm{C,J})}(y) \equiv -\sqrt{1-y^2}\,\frac{\mathrm{d}}{\mathrm{d}y}\ln\left|\psi_0^{(\mathrm{C,J})}(x)\right| ~,
\label{eq:SWKB_CESJ}
\end{multline}
where $a'_{\mathrm{L}},a'_{\mathrm{R}}~(a'_{\mathrm{L}} < a'_{\mathrm{R}})$ are the two solutions for the equation obtained by setting the inside of the square root equals zero.
The SWKB condition is now
\begin{equation}
I^{(\mathrm{C},\ast)} = n\pi ~,~~~
n \in \mathbb{Z}_{\geqslant 0} ~.  
\label{eq:SWKBcond_CES}
\end{equation}
The SWKB conditions \eqref{eq:SWKB_CESH}--\eqref{eq:SWKB_CESJ} are totally independent of $\hbar$ (and $\omega$), which means that this condition equation is not to be discussed in the context of the semi-classical regime of the quantum system.

\subsubsection{Comparison of pole structures}

\begin{figure}
\centering
	\begin{minipage}[t]{0.49\textwidth}
	\centering
	\subfloat[][$b=-0.5$.]{
	\includegraphics[scale=1]{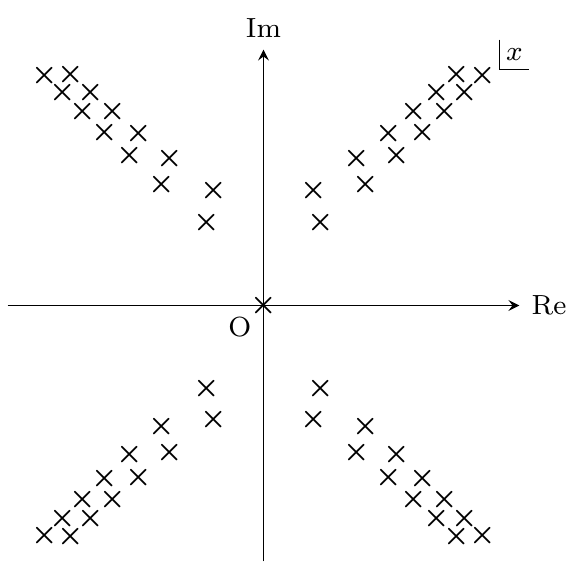}
	\label{}
	}
	\end{minipage}
	\begin{minipage}[t]{0.49\textwidth}
	\centering
	\subfloat[][$b=0$.]{
	\includegraphics[scale=1]{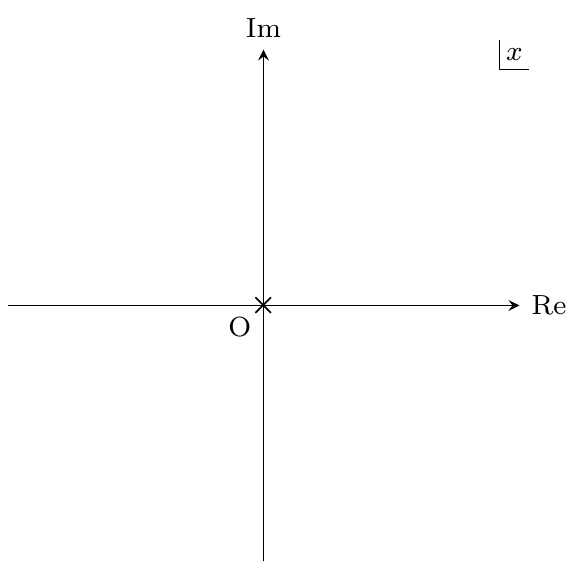}
	\label{fig:SingularityStructs_CESH0}
	}
	\end{minipage}
	
	\bigskip
	\begin{minipage}[t]{0.49\textwidth}
	\centering
	\subfloat[][$b=0.1$.]{
	\includegraphics[scale=1]{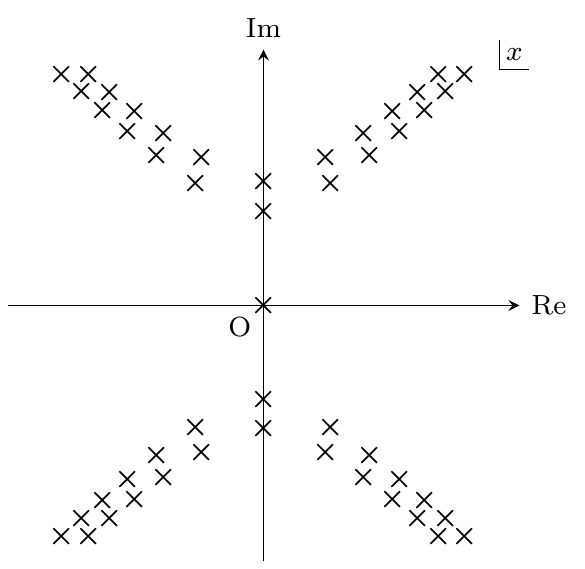}
	\label{fig:SingularityStructs_CESH01}
	}
	\end{minipage}
	\begin{minipage}[t]{0.49\textwidth}
	\centering
	\subfloat[][$b=3.5$.]{
	\includegraphics[scale=1]{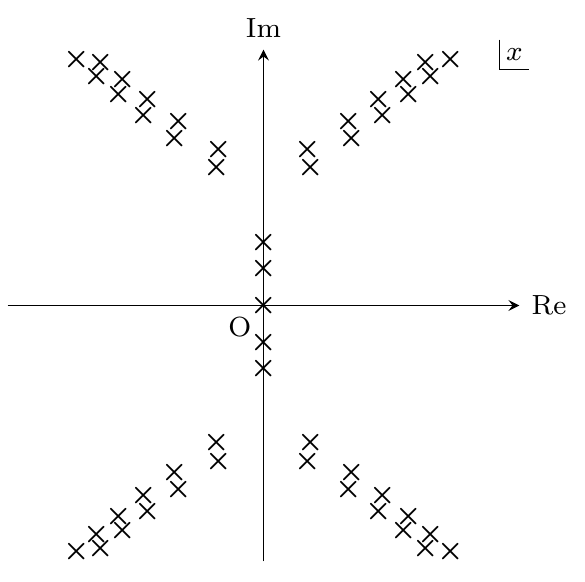}
	\label{fig:SingularityStructs_CESH35}
	}
	\end{minipage}
	
	\bigskip
	\begin{minipage}[t]{0.49\textwidth}
	\centering
	\subfloat[][$b=4$.]{
	\includegraphics[scale=1]{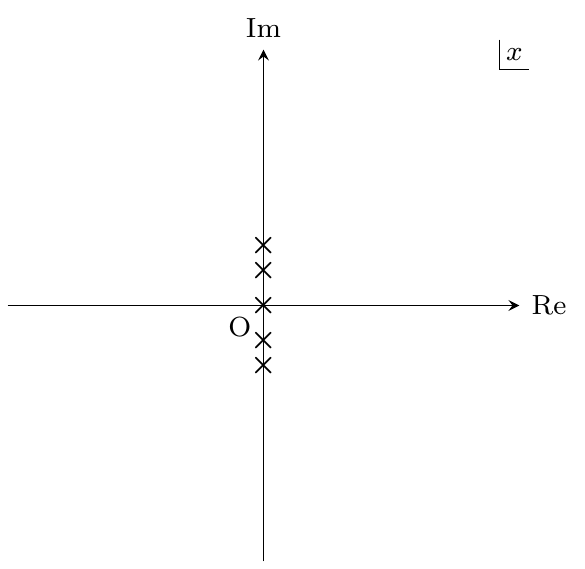}
	\label{fig:SingularityStructs_CESH4}
	}
	\end{minipage}
	\begin{minipage}[t]{0.49\textwidth}
	\centering
	\subfloat[][$b=4.1$.]{
	\includegraphics[scale=1]{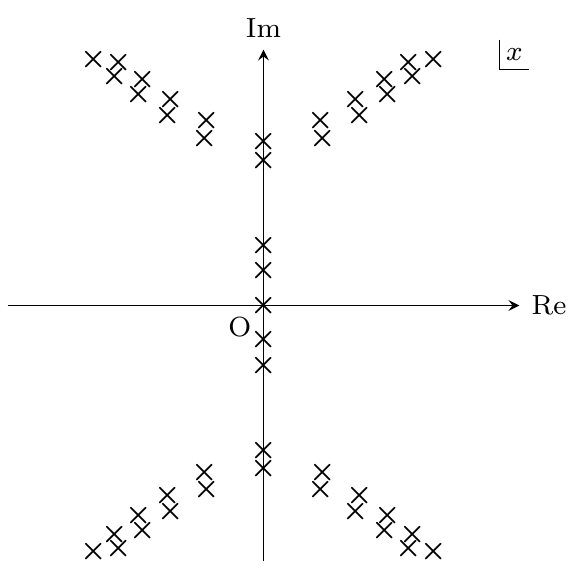}
	\label{fig:SingularityStructs_CESH41}
	}
	\end{minipage}
\caption{The singularity structures of the QMFs for the first excited states of the CES (H) systems with various $b$.
	Poles are plotted by x-marks.
	The location of each pole is calculated numerically.
	Note that (b) and (e) are identical to that of the 1-d H.O. and the Krein--Adler (H) system with $d=1$ respectively.}
\label{fig:SingularityStructs_CESH}
\end{figure}

\begin{figure}[t]
\centering
\includegraphics[scale=1]{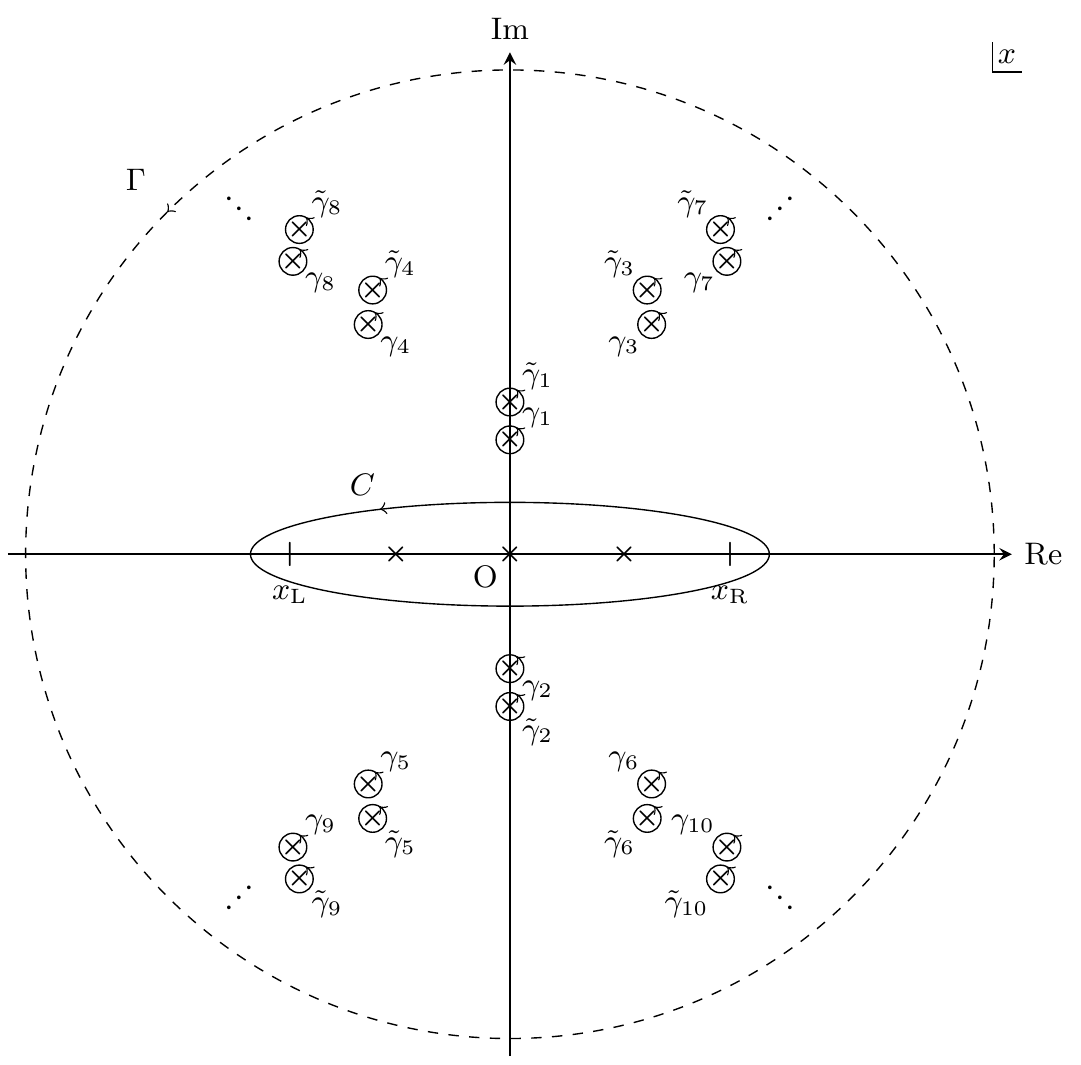}
\caption{The contours of the integrations in Eq.~\eqref{eq:contours_rel_CESH}.
	The dots $\cdots$ show that the infinite number of pairwise poles lie in the plane.
	The dashed contour $\Gamma$ is a virtual contour that would enclose all the poles except for the one at infinity.}
\label{fig:contours_CES}
\end{figure}

\begin{figure}
\centering
	\begin{minipage}[t]{0.49\textwidth}
	\centering
	\subfloat[][$b=-0.5$.]{
	\includegraphics[scale=1]{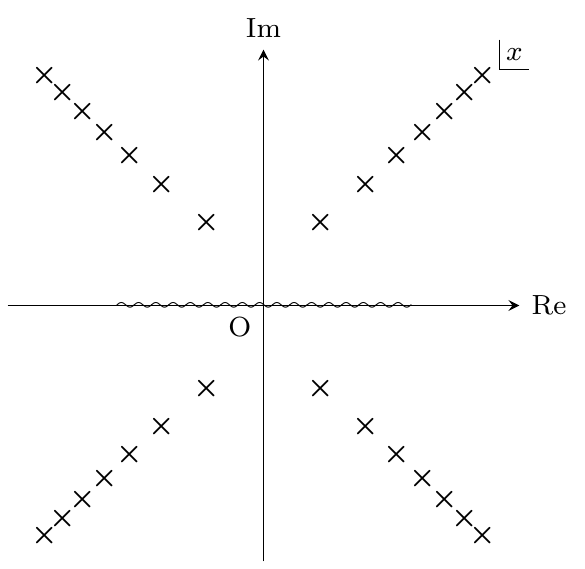}
	\label{fig:SingularityStructs_CESHm05_SWKB}
	}
	\end{minipage}
	\begin{minipage}[t]{0.49\textwidth}
	\centering
	\subfloat[][$b=0$.]{
	\includegraphics[scale=1]{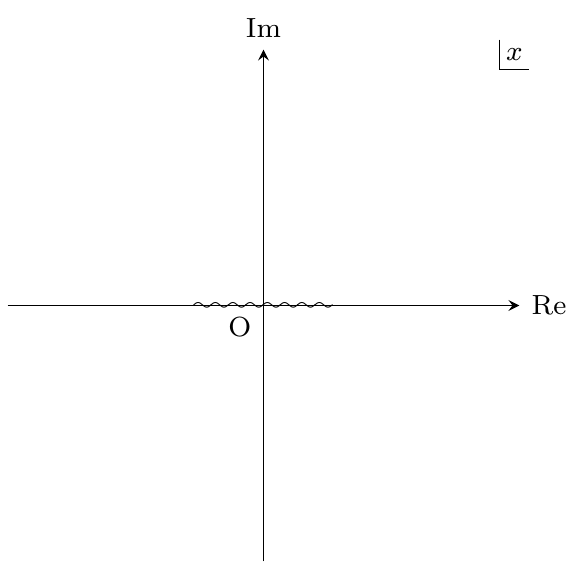}
	\label{fig:SingularityStructs_CESH0_SWKB}
	}
	\end{minipage}
	
	\bigskip
	\begin{minipage}[t]{0.49\textwidth}
	\centering
	\subfloat[][$b=0.1$.]{
	\includegraphics[scale=1]{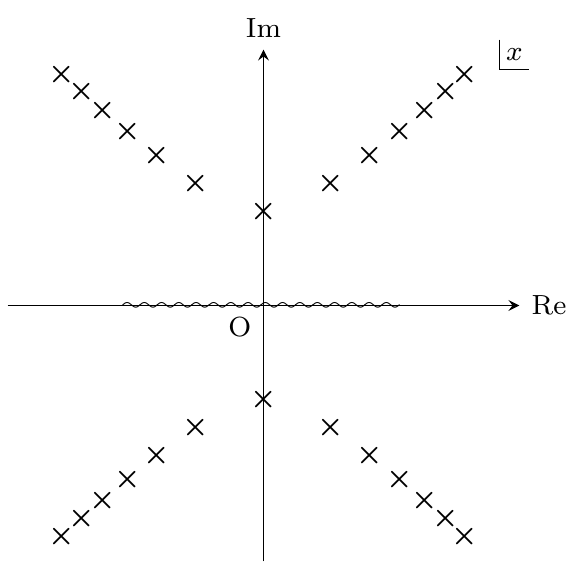}
	\label{fig:SingularityStructs_CESH01_SWKB}
	}
	\end{minipage}
	\begin{minipage}[t]{0.49\textwidth}
	\centering
	\subfloat[][$b=3.5$.]{
	\includegraphics[scale=1]{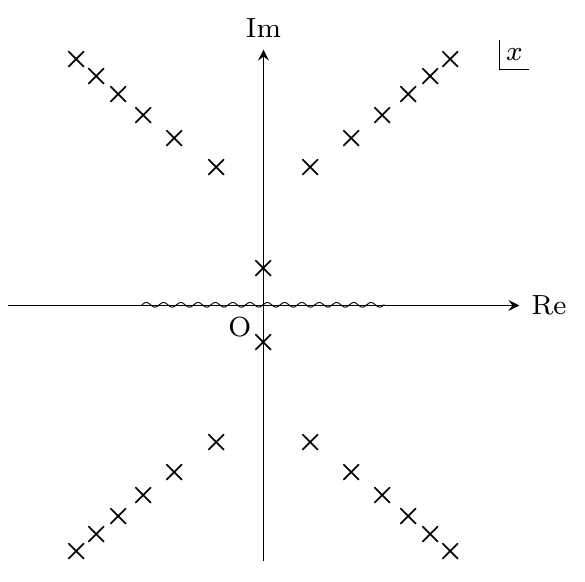}
	\label{fig:SingularityStructs_CESH35_SWKB}
	}
	\end{minipage}
	
	\bigskip
	\begin{minipage}[t]{0.49\textwidth}
	\centering
	\subfloat[][$b=4$.]{
	\includegraphics[scale=1]{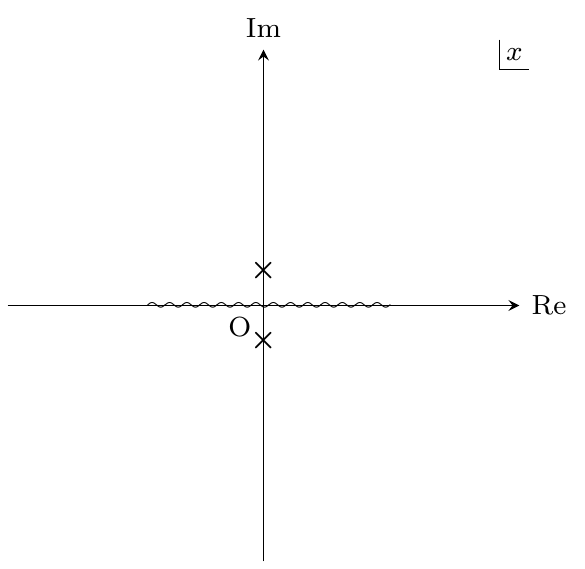}
	\label{fig:SingularityStructs_CESH4_SWKB}
	}
	\end{minipage}
	\begin{minipage}[t]{0.49\textwidth}
	\centering
	\subfloat[][$b=4.1$.]{
	\includegraphics[scale=1]{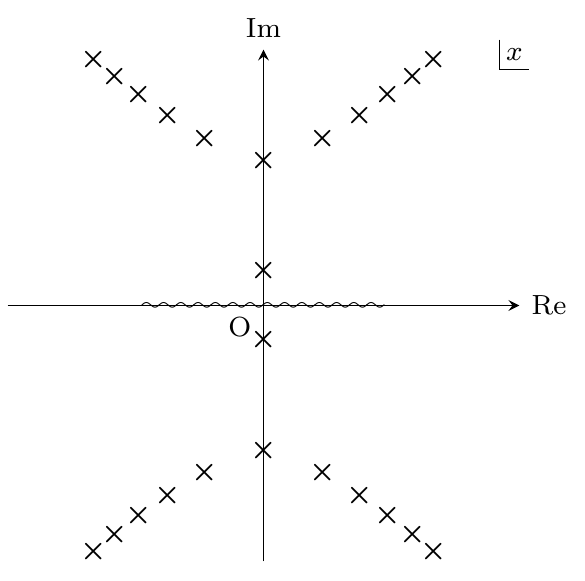}
	\label{fig:SingularityStructs_CESH41_SWKB}
	}
	\end{minipage}
\caption{The singularity structures of the SWKB integrands for the first excited states of the CES (H) systems with various $b$.
	The poles are plotted by x-marks and the branch cut on the real axis is shown by wavy lines.
	Other branch cuts are removed from these cartoons.
	The location of each pole is calculated numerically.
	Note that (b) and (e) are identical to that of the 1-d H.O. and the Krein--Adler (H) system with $d=1$ respectively.}
\label{fig:SingularityStructs_CESH_SWKB}
\end{figure}

We compare the singularity structures of both the SWKB integrand and the QMF.
In this subsection, we fix $\hbar=\omega=1$ without loss of generality.
First, the pole structures of the QMF:
\begin{equation}
p(x;\mathcal{E}_n) = -\mathrm{i}\hbar\left[ \frac{\psi_0^{(\mathrm{C},\ast)'}(x)}{\psi_0^{(\mathrm{C},\ast)}(x)} + \frac{u^{(\ast)}(x)\psi_n^{(\mathrm{C},\ast)'}(x)}{cu^{(\ast)}(x)\phi_n^{(\ast)}(x) + u^{(\ast)'}(x)\psi_{n-1}^{(\ast,+)}(x)} + \frac{u^{(\ast)'}(x)}{u^{(\ast)}(x)} \right] ~,
\end{equation}
with $c$ being some constant, are presented in Fig.~\ref{fig:SingularityStructs_CESH}.
Note that these figures show our numerical results with several $b$, where we set $\ast=\mathrm{H}$, $\beta=0$ and $n=1$.
They reveal notable features of the QMF of the CES systems.
Except for $b=0$ (conventional SI) and $b=4N$ (Krein--Adler), the QMF has infinite number of poles in the complex plane.
At $b=0$, there is just one pole at the origin $x=0$ (Fig.~\ref{fig:SingularityStructs_CESH0}).
For $b\neq 0$ , infinite number of poles appear in the complex plane (Fig.~\ref{fig:SingularityStructs_CESH01}) and also 
$4N+1$ poles on the imaginary axis for $4(N-1) < b \leqslant 4N$.
The poles on the imaginary axis approach the origin $x=0$ as $b$ grows, while the other poles remain almost the same locations 
(Fig.~\ref{fig:SingularityStructs_CESH35}).
When $b$ reaches $4N$, all the poles except the ones on the imaginary axis disappear (Fig.~\ref{fig:SingularityStructs_CESH4}).
Again, as $b$ grows further, infinite poles appear in the complex plane (Fig.~\ref{fig:SingularityStructs_CESH41}). 
A notable feature is that these poles except for the origin $x=0$ (and the one at $x\to\infty$) are pairwise 
with the residues $1$ and $-1$, respectively.
Therefore, for the contour integral of $J_{\mathrm{QHJ}}$, these contributions exactly vanish and only the 
residue at $x\to\infty$ contributes to the integral;
\begin{equation}
J_{\Gamma} = J_{\mathrm{QHJ}} + \sum_{i=1}^{\infty} J_{\gamma_i} + \sum_{j=1}^{\infty} J_{\tilde{\gamma}_j} 
= J_{\mathrm{QHJ}} ~,
\label{eq:contours_rel_CESH}
\end{equation}
which we have numerically verified.
For the definitions of the contours, see Fig.~\ref{fig:contours_CES}.
Note that this is just the quantization of the quantum action variable, not the quantization of the energy, and then there is no direct method for calculating the energy $\mathcal{E}$ from the quantization condition.

For the SWKB integration, the situation is worse. 
Infinite number of the poles appeared in the complex plane are not pairwise and then, no cancellation of residue of the poles occurs (see Fig.~\ref{fig:SingularityStructs_CESH_SWKB}). 
Also, there appear branch cuts other than the one on the real axis (which are sometimes referred to as ``other branch cuts'').
They have nonzero contribution on the contour integral $J_{\mathrm{SWKB}}$.
They spread all over the complex plane, but we do not plot in Fig.~\ref{fig:SingularityStructs_CESH_SWKB} for making easier to see.
These are an origin of the non-exactness of the SWKB condition and also the essential difficulty 
for the explicit calculation of $J_{\mathrm{SWKB}}$.
This gives us an intuition that we have to rely on a perturbative treatment to analyze the condition further.

As a result, both $J_{\mathrm{QHJ}}$ and $J_{\mathrm{SWKB}}$ cannot analytically reproduce the energy spectra.  
The SWKB formalism has some advantages because the original $I_{\mathrm{SWKB}}$ can be integrated along the real axis, without summing up all the residues of the poles in the complex plane.
Therefore, if we successfully prove that a system satisfies the SWKB condition within some uncertainty, we are able to compute all the energy spectra approximately.

\subsubsection{Numerical results}

\begin{figure}
\centering
\hspace*{-.05\linewidth}
	\begin{minipage}[t]{0.45\linewidth}
	\centering
        \subfloat[][$\beta=0$.]{
          \includegraphics[scale=0.63]{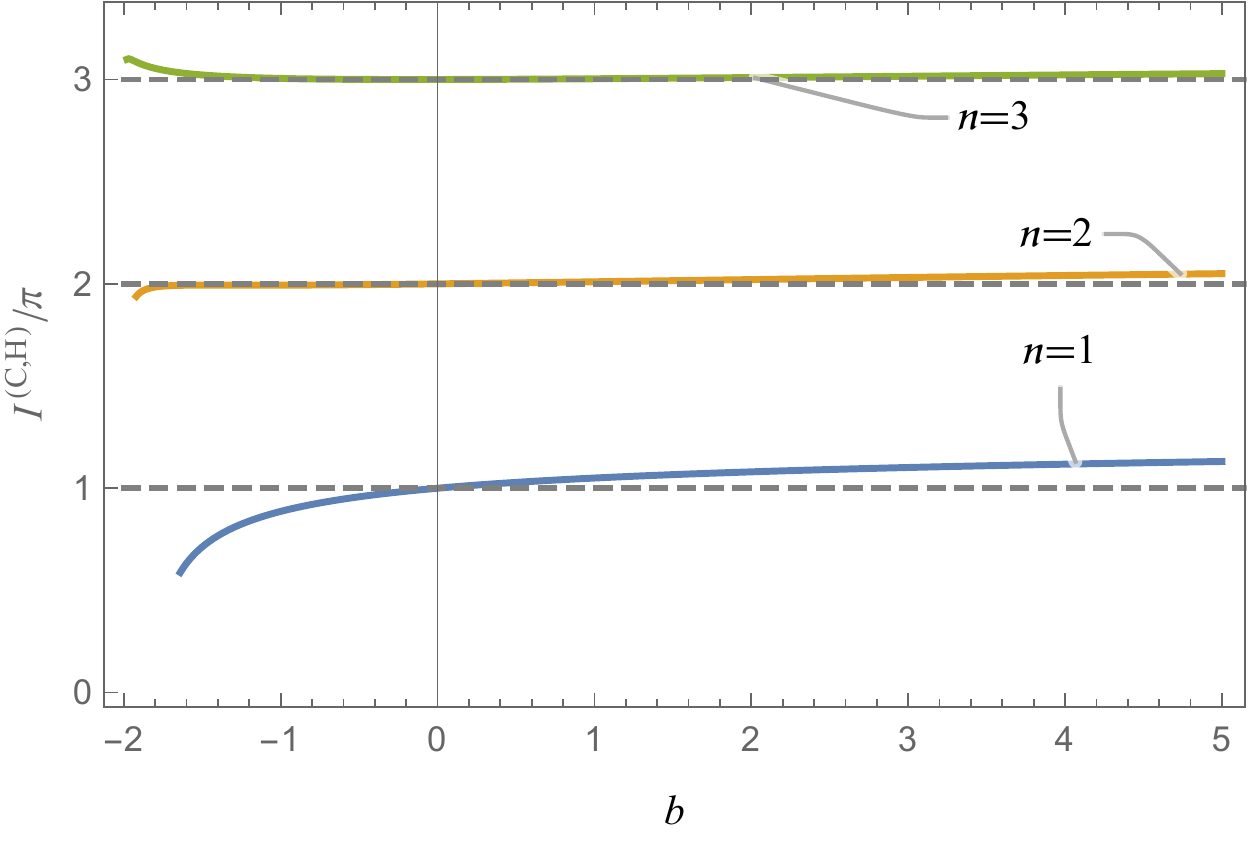}
          \label{fig:SWKB_CESH_b}
        }
	\end{minipage}
\qquad\quad
	\begin{minipage}[t]{0.45\linewidth}
	\centering
        \subfloat[][$b=0$.]{
          \includegraphics[scale=0.63]{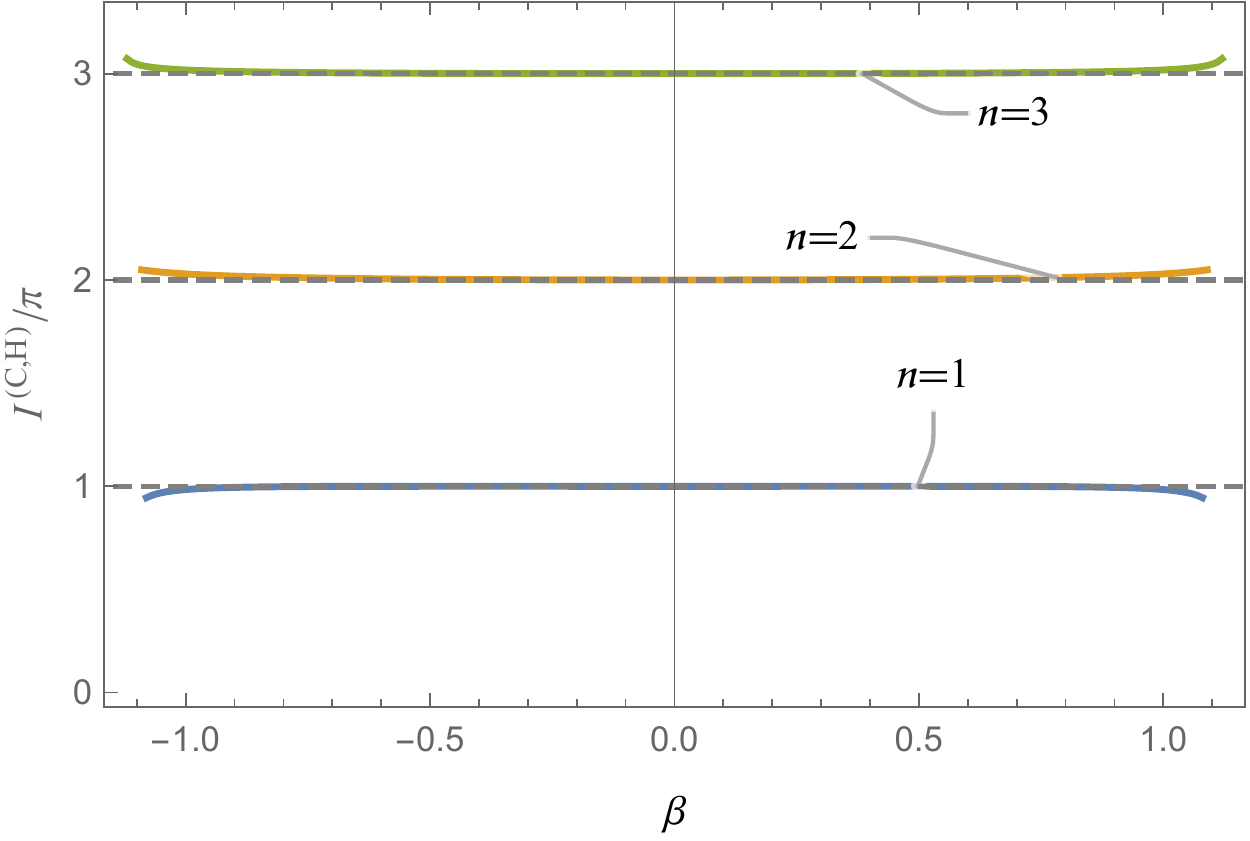}
          \label{fig:SWKB_CESH_beta}
        }
        \end{minipage}
\caption{The values of the SWKB integral $I^{(\mathrm{C,H})}$ for $n=1,2,3$.
	Range of each plot is determined so that the systems will not have more than two turning points.
	The parametric conditions \eqref{eq:cond1_CESH} and \eqref{eq:cond2_CESH} yield $b>-2$ and $|\beta| < 2/\sqrt{\pi}$ respectively.}
\label{fig:SWKB_CESH}
\end{figure}

The comparison of Fig.~\ref{fig:SingularityStructs_CESH_SWKB} with Fig.~\ref{fig:SingularityStructs_CESH} indicates that the SWKB condition \eqref{eq:SWKBcond_CES} breaks, which we demonstrate numerically.
Fig.~\ref{fig:SWKB_CESH_b} shows the $b$-dependency of the SWKB integral $I^{(\mathrm{C,H})}$ with $\beta=0$, and Fig.~\ref{fig:SWKB_CESH_beta} is the $\beta$-dependency of $I^{(\mathrm{C,H})}$ with $b=0$.
The SWKB integrals $I^{(\mathrm{C,H})}$ grow with the parameter $b$ around $b=0$, while $I^{(\mathrm{C,H})}$ exhibit plateau behavior (but the condition is never exactly satisfied) around $\beta=0$.
Different behaviors are seen as the parameters approach their boundary \eqref{eq:cond1_CESH},\eqref{eq:cond2_CESH}.
These statements hold for general $b\neq 0$ and $\beta \neq 0$ case.
Similar results can be obtained for the cases of $\ast = \mathrm{L,J}$.
The numerical calculations Fig.~\ref{fig:SWKB_CESH} support our conjecture~\cite{nasuda2021numerical} that the level structure approximately guarantees the exactness of the SWKB condition.

In order to calculate exact bound-state energy spectra through the SWKB condition, one needs to evaluate $\Delta = n\pi\hbar - I_{\mathrm{SWKB}}$. 
Since the quantization condition for the quantum action variable gives exact results, one may think that all one has to do is to evaluate $J_{\mathrm{QHJ}}$.
However, in general, one cannot calculate the analytical relation between $J_{\mathrm{QHJ}}$ and the energy because of the complicated singularity structures of the QMF.
In the latter half of Sec.~\ref{sec:Delta}, we propose a noble method for evaluating $\Delta$ by means of series expansion of the SWKB integrand.

\subsection{SWKB condition for Krein--Adler systems}

Before we analyze $\Delta$, we see the case of $b=4N$ and $\beta=0$, \textit{i.e.}, Krein--Adler systems,  where one can partly carry out analytical calculation for the discussions above.
The SWKB integral \eqref{eq:SWKB} for the Krein--Adler systems is
\begin{equation}
I_{\mathrm{SWKB}} =
\int_{a_{\mathrm{L}}}^{a_{\mathrm{R}}} \sqrt{\mathcal{E}_{\mathcal{D};n}^{\mathrm{(K,\ast)}} - \left( \hbar\frac{\mathrm{d}}{\mathrm{d}x}\ln\left|\psi_{\mathcal{D};0}^{\mathrm{(K,\ast)}}(x)\right| \right)^2}\,\mathrm{d} x ~.
\label{eq:SWKB_KA}
\end{equation}
For the case of $\ast=\mathrm{H}$, Eq.~\eqref{eq:SWKB_KA} reduces to
\begin{equation}
I_{\mathrm{SWKB}} 
= \hbar\int_{a'_{\mathrm{L}}}^{a'_{\mathrm{R}}} \sqrt{2\breve{n} - \left( \frac{\mathrm{d}}{\mathrm{d}\xi}\ln\left|\psi_{\mathcal{D};0}^{\mathrm{(K,H)}}(x)\right| \right)^2}\,\mathrm{d}\xi 
\equiv \hbar I^{(\mathrm{K,H})} ~,
\label{eq:SWKB_KAH}
\end{equation}
while for the cases of $\ast=\mathrm{L, J}$, Eq.~\eqref{eq:SWKB_KA} becomes
\begin{gather}
I_{\mathrm{SWKB}} 
= \hbar\int_{a'_{\mathrm{L}}}^{a'_{\mathrm{R}}} \sqrt{\breve{n} - z\left( \frac{\mathrm{d}}{\mathrm{d}z}\ln\left|\psi_{\mathcal{D};0}^{\mathrm{(K,L)}}(x)\right| \right)^2}\,\frac{\mathrm{d}z}{\sqrt{z}} 
\equiv \hbar I^{(\mathrm{K,L})} ~, 
\label{eq:SWKB_KAL} \\
I_{\mathrm{SWKB}}
= \hbar\int_{a'_{\mathrm{L}}}^{a'_{\mathrm{R}}} \sqrt{\breve{n}(\breve{n}+g+h) - \left( 1-y^2 \right)\left( \frac{\mathrm{d}}{\mathrm{d}y}\ln\left|\psi_{\mathcal{D};0}^{(\mathrm{K,J})}(x)\right| \right)^2}\,\frac{\mathrm{d}y}{\sqrt{1-y^2}} 
\equiv \hbar I^{(\mathrm{K,J})} ~.
\label{eq:SWKB_KAJ}
\end{gather}
For each case above, $a'_{\mathrm{L}},a'_{\mathrm{R}}~(a'_{\mathrm{L}} < a'_{\mathrm{R}})$ denote the two solutions of the equation obtained by setting the inside of the square root equals zero.
The SWKB condition is then
\begin{equation}
I^{(\mathrm{K},\ast)} = n\pi ~,~~~
n \in \mathbb{Z}_{\geqslant 0} ~.  
\end{equation}
Since the SWKB conditions \eqref{eq:SWKB_KAH}--\eqref{eq:SWKB_KAJ} are totally independent of $\hbar$ (and $\omega$)~\cite{nasuda2021numerical}, 
for the rest of this section we fix $\hbar=\omega=1$ without loss of generality.

\begin{figure}
	\begin{minipage}[t]{.4\linewidth}
	\centering
        \subfloat[][The QMF.]{
          \includegraphics[scale=1.5]{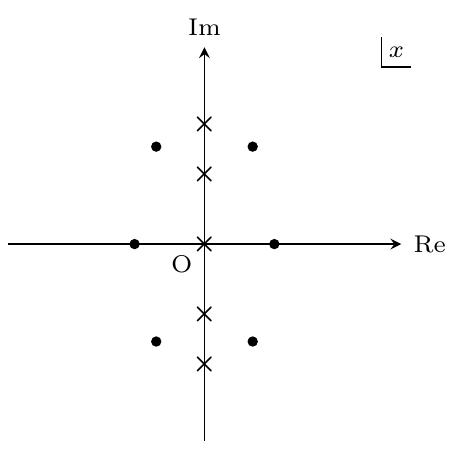}
		}
	\end{minipage}
\qquad\qquad
	\begin{minipage}[t]{.4\linewidth}
	\centering
        \subfloat[][The SWKB integrand.]{
	  \includegraphics[scale=1.5]{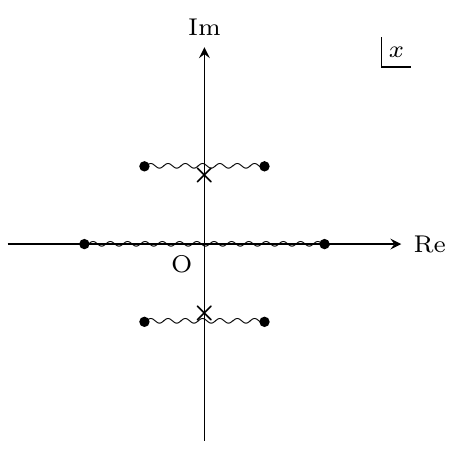}
        }
	\end{minipage}
\caption{Poles (x-marks) and branch cuts (wavy lines) on a complex $x$-plane are displayed for (a) the QMF $p(x;\mathcal{E}_1)$ for a Krein--Adler system, and (b) the SWKB integrand $\sqrt{\mathcal{E}_1-W(x)^2}$.
	We also plot the zeros with closed dots on both figures.
	The poles are at (a) $x=0,~\pm\mathrm{i}/\sqrt{2},~\pm\mathrm{i}\sqrt{3/2}$, (b) $x=\pm\mathrm{i}/\sqrt{2}$ and the nodes exist at (a) $x=\pm1/\sqrt{2},~\left( \pm\sqrt[4]{33-12\sqrt{6}} \pm\sqrt[4]{33+12\sqrt{6}} \right) \Big/ 2\sqrt{2},
	~\left( \pm\sqrt[4]{33-12\sqrt{6}} \mp\sqrt[4]{33+12\sqrt{6}} \right) \Big/ 2\sqrt{2}$, (b) $x=\pm\sqrt{3/2},~\left( \pm\sqrt{3}\pm\mathrm{i}\sqrt{5} \right) \big/ 2\sqrt{2}$.}
\label{fig:Poles_BranchCuts}
\end{figure}

As in the previous subsection, we compare the singularity structures of the QMF and the SWKB integrand.
Here, we choose $\ast = \mathrm{H}$ and $\mathcal{D}=\{ d,d+1 \}$ as an example.
The QMF of the system is
\begin{align}
p(x,\mathcal{E}_n) &= -\mathrm{i} \left( -x - \frac{\mathrm{d}}{\mathrm{d}x}\ln\mathrm{W}[H_d,H_{d+1}](x) + \frac{\mathrm{d}}{\mathrm{d}x}\ln\mathrm{W}[H_d,H_{d+1},H_{\breve{n}}](x) \right) \nonumber \\
&= -\mathrm{i} \left( \frac{\psi_{\mathcal{D};0}^{(\mathrm{K,H})'}(x)}{\psi_{\mathcal{D};0}^{(\mathrm{K,H})}(x)} - \frac{\mathrm{d}}{\mathrm{d}x}\ln\mathrm{W}[H_d,H_{d+1},1](x) + \frac{\mathrm{d}}{\mathrm{d}x}\ln\mathrm{W}[H_d,H_{d+1},H_{\breve{n}}](x) \right) ~.
\end{align}
The plots of the singularity structures for both the QMF and the SWKB integrand for the first excited state with $d = 1$ are displayed in Fig.~\ref{fig:Poles_BranchCuts}.
These figures are the results of the analytic calculation; the position of each singularity is obtained analytically.
Apparently they do not coincide with each other and the quantization of the SWKB integral is not exact.

\begin{figure}
\centering
\includegraphics[scale=1]{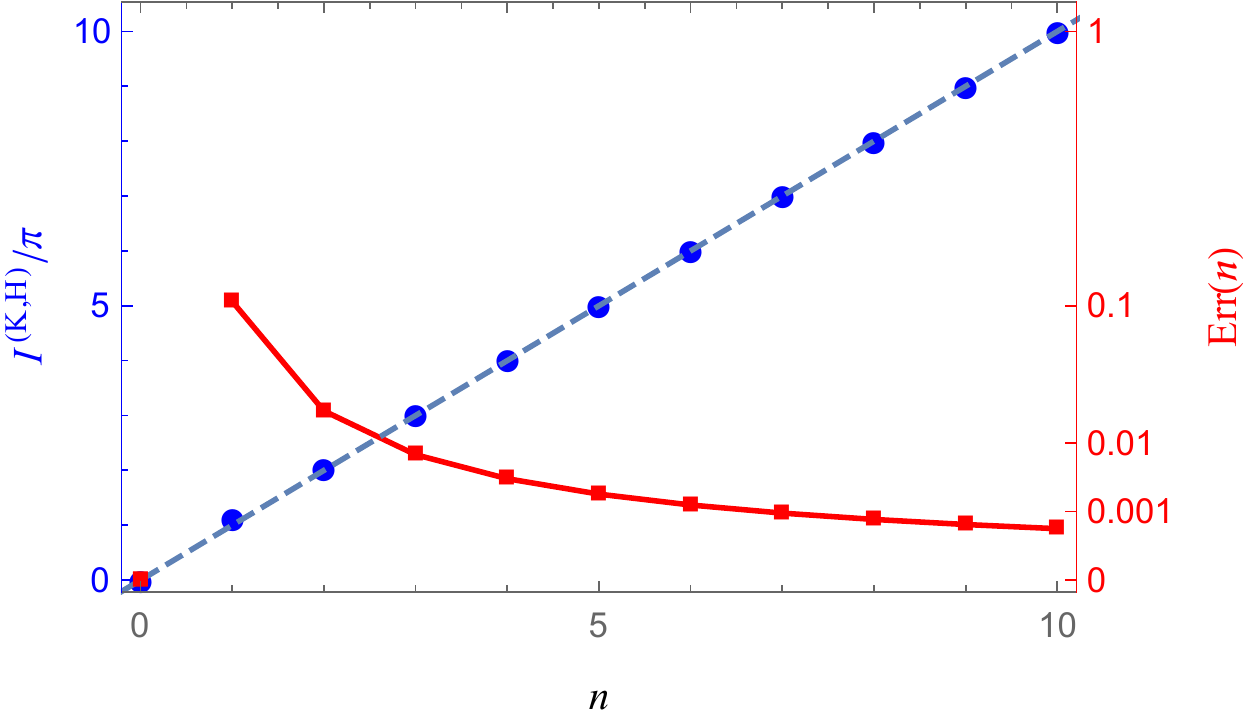}
\caption{The values of the SWKB integral $I^{(\mathrm{K,H})}$ (blue closed dots) and the accuracy of the SWKB condition $\mathrm{Err}(n)$ (red squares) for the case $d=1$ are plotted.
	The condition is exact when the blue dots are on the blue dashed line $I^{(\mathrm{K,H})}/\pi = n$.
	The maximal error is $1.1\times 10^{-1}$, which is found at $n=1$.}
\label{fig:SWKB_KA}
\end{figure}

We show in Fig.~\ref{fig:SWKB_KA} the breaking of the SWKB condition directly by the numerical evaluations of $I^{(\mathrm{K,H})}$.
We also display in Fig.~\ref{fig:SWKB_KA} the accuracy of the SWKB conditions calculated by
\begin{equation}
\mathrm{Err}(n) \coloneqq \frac{I^{(\mathrm{K,H})} - n\pi}{I^{(\mathrm{K,H})}} ~.
\end{equation}
As expected, though the SWKB condition is not exact except for the ground state, the error $\mathrm{Err}(n)$ remains small, at most $\mathrm{Err}(n) \sim 10^{-1}$.
$\mathrm{Err}(n)$ has its maximal value at $n=1$, and as $n$ gets larger, $\mathrm{Err}(n)$ monotonically decay; 
$\mathrm{Err}(n)$ goes to zero as $n\to\infty$.
We thus conclude that the SWKB condition is not exact but still reproduces approximate bound-state spectra for these cases.
The similar can be said for other choices of parameters.
The maximal value of $\mathrm{Err}(n)$ is seen around $n=d$.
More comprehensive results are given in~\cite{nasuda2021numerical}.

\begin{figure}
	\begin{minipage}[b]{.4\linewidth}
	\centering
        \subfloat[][$x$-plane.]{
	\includegraphics[scale=1.25]{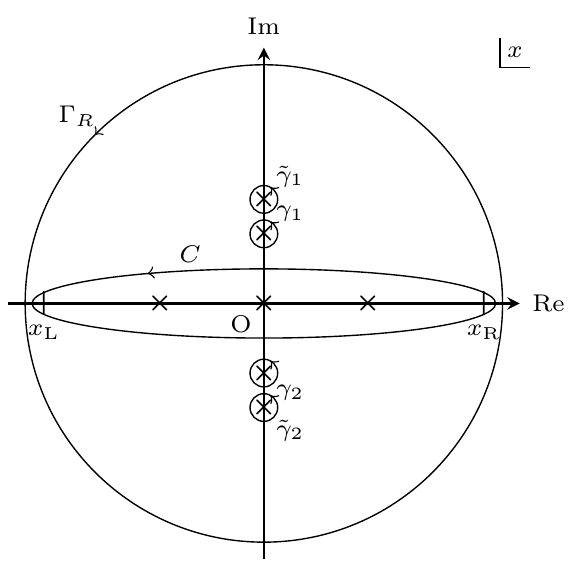}
	\label{fig:contours_KAa}
        }
	\end{minipage}
\qquad\qquad
	\begin{minipage}[b]{.4\linewidth}
	\centering
        \subfloat[][$w$-plane.]{
          \includegraphics[scale=1.5]{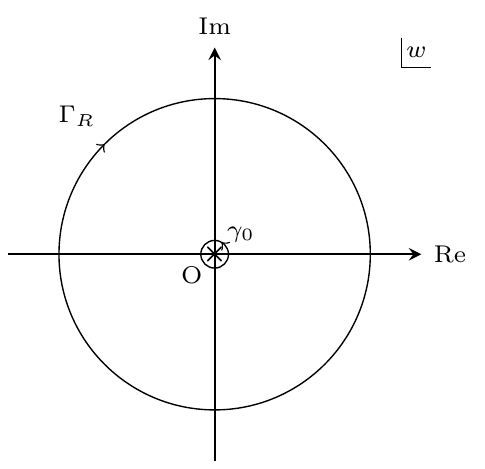}
          \label{fig:contours_KAb}
	}
	\end{minipage}
\caption{The contours of the integrations in (a) Eq.~\eqref{eq:contours_rel} and (b) Eq.~\eqref{eq:J_QHJ_XtoW_KA}.}
\label{fig:contours_KA}
\end{figure}

As is well-known in the literature, the quantization condition for the quantum action variable is always exact, which is proved by the Cauchy's argument principle and the node theorem.
Furthermore for this class of solvable systems, the analytical contour integrations for the poles are executable.
This enables us to obtain the energy spectrum analytically.
As we mentioned, $C$ in Eq.~\eqref{eq:QHJ} is the counterclockwise contour enclosing the two classical turning points $x_{\mathrm{L,R}}$.
For the Krein--Adler system, the QMF has an isolated pole at $x\to\infty$, $4d-2$ fixed poles other than that and $\breve{n}$ moving poles, including $n$ moving poles on the real axis.
The contour $\Gamma_R$ is of the radius $R$, enclosing all $4d-2$ fixed poles counterclockwise, where each of the poles is enclosed by a counterclockwise contour $\gamma_i$.
$\breve{n}-n$ moving poles, which are off the real axis, are enclosed by a counterclockwise contour $\tilde{\gamma}_j$ one by one.
See Fig.~\ref{fig:contours_KAa}.
Hence, the following equation holds:
\begin{equation}
J_{\Gamma_R} = J_{\mathrm{QHJ}} + \sum_{i=1}^{4d-2}J_{\gamma_i} + \sum_{j=1}^{\breve{n}-n}J_{\tilde{\gamma}_j} ~,
\label{eq:contours_rel}
\end{equation}
where $J_{\bullet}$ is defined as
\begin{equation}
J_{\bullet} \coloneqq \frac{1}{2\pi} \oint_{\bullet} p(x;\mathcal{E}) \,\mathrm{d}x ~.
\end{equation}

Here, considering $\mathrm{W}[H_d,H_{d+1}](x)$, $\mathrm{W}[H_d,H_{d+1},1](x)$ and $\mathrm{W}[H_d,H_{d+1},H_{\breve{n}}](x)$ are polynomials of $2d$, $2d-2$ and $2d-2+\breve{n}$ degrees respectively, the second and the third terms of the r.h.s. of Eq.~\eqref{eq:contours_rel} are 
\begin{align}
\sum_{i=1}^{4d-2}J_{\gamma_i} &= (2d-2) - 2d = -2 ~, \\
\sum_{j=1}^{\breve{n}-n}J_{\tilde{\gamma}_j} &= -(2d-2) + (2d-2+\breve{n}-n) = \breve{n}-n ~.
\end{align}

In order to evaluate $J_{\Gamma_R}$, we change variable as $x \to w \equiv x^{-1}$, and
\begin{equation}
J_{\Gamma_R} = \frac{1}{2\pi} \oint_{\Gamma_R} p(x;\mathcal{E}) \,\mathrm{d}x
= \frac{1}{2\pi} \oint_{\gamma_0} \bar{p}(w;\mathcal{E}) \,\frac{\mathrm{d}w}{w^2} ~,
\label{eq:J_QHJ_XtoW_KA}
\end{equation}
with $\bar{p}(w;\mathcal{E}) \equiv p(w^{-1};\mathcal{E})$ and $\gamma_0$ enclosing counterclockwise the only pole in the $w$-plane, \textit{i.e.}, the one at $w=0$ (see Fig.~\ref{fig:contours_KAb}).
Note that $\bar{p}(w;\mathcal{E})$ satisfies the QHJ equation:
\begin{equation}
\bar{p}(w;\mathcal{E})^2 + \mathrm{i} w^2 \frac{\mathrm{d}\bar{p}(w;\mathcal{E})}{\mathrm{d}w} = \mathcal{E} - \bar{V}^{(\mathrm{K,H})}(w) ~.
\label{eq:QHJeq_y}
\end{equation}
Here, $\bar{V}^{(\mathrm{K,H})}(w)$ means the potential for the Krein--Adler system with $\ast=\mathrm{H}$ in terms of the variable $w$.

We employ the Laurent expansion of $\bar{p}(w;\mathcal{E})$ about $w=0$:
\begin{equation}
\bar{p}(w;\mathcal{E}) \cong \sum_{n=0}^{\infty} a_nw^n + \sum_{q=1}^{k} \frac{b_q}{w^q}
\end{equation}
to calculate $J_{\Gamma_R}$.
Substituting this expansion into Eq.~\eqref{eq:QHJeq_y} and comparing the l.h.s. and the r.h.s., one obtains
\begin{align}
&{a_0}^2 + 2a_1b_1 - \mathrm{i}b_1 = \mathcal{E} - 3 ~, \\
&{b_1}^2 = -1 ~, \\
&2a_0b_1 = 0 ~,
\end{align}
and $b_q=0$ for $q \geqslant 2$.
The asymptotic behavior of $\bar{p}$ leads $b_1 = \mathrm{i}$, and $a_0=0$.
Hence, $J_{\Gamma_R} = \mathcal{E}/2 - 2$, and the quantization condition yields
\begin{equation}
\mathcal{E} = 2\breve{n} ~.
\end{equation}

On the other hand, the contour integral for $J_{\mathrm{SWKB}}$ is not so straightforward. 
The contour integrations for the singularities cannot be performed analytically. 
The different number of the poles and the existence of other branch cuts cause the difficulty, and also are responsible for the breaking. 
We quantitatively confirmed the explicit reasons for the broken quantization condition of SWKB, but no way of calculating it exactly. 
Instead, a perturbative treatment works for the discrepancy $\Delta = \pi J_{\mathrm{QHJ}} - I_{\mathrm{SWKB}}$, which we see in the next section. 
An analytical derivation of the energy eigenvalues from the SWKB formalism is presented in Appendix~\ref{sec:Appx} for a simplest case.

\section{Analysis on the residual}
\label{sec:Delta}

\subsection{Series expansion of SWKB integrand}

We first investigate how the SWKB integrals $I^{(\mathrm{C},\ast)}$ change as the parameters $b,\beta$ grow by a series expansion of the SWKB integrand.
Note that for $b=\beta=0$, the condition becomes exact, since the systems are equivalent to the original conventional SI ones.
Also, it is notable that for the exact case the main part of the SWKB integral is of the form $\sqrt{(x-a_{\mathrm{L}})(a_{\mathrm{R}}-x)}$.

Our basic idea for the formulation is to consider small perturbations from the exact case: $b=\beta=0$.
We employ Taylor expansion for the SWKB integrand around the point where the SWKB condition is exact.
For the case of $\ast=\mathrm{H}$,
\begin{align}
I^{(\mathrm{C,H})} 
&= \int_{a_{\mathrm{L}}}^{a_{\mathrm{R}}} \sqrt{(x-a_{\mathrm{L}})(a_{\mathrm{R}}-x)}\sqrt{1+\frac{2n + b - \widetilde{W}^{(\mathrm{C,H})}(x)^2 - (x-a_{\mathrm{L}})(a_{\mathrm{R}}-x)}{(x-a_{\mathrm{L}})(a_{\mathrm{R}}-x)}}\,\mathrm{d}x \nonumber \\
&\cong \frac{(a_{\mathrm{R}} - a_{\mathrm{L}})^2}{8}\pi + \sum_{k=1}^{\infty}\frac{(-1)^k(2k)!}{(1-2k)(k!)^24^k} \int_{a_{\mathrm{L}}}^{a_{\mathrm{R}}} \frac{ \left[ 2n + b - \widetilde{W}^{(\mathrm{C,H})}(x)^2 - (x-a_{\mathrm{L}})(a_{\mathrm{R}}-x) \right]^k}{\left[ (x-a_{\mathrm{L}})(a_{\mathrm{R}}-x) \right]^{k-\frac{1}{2}}}\,\mathrm{d}x ~.
\label{eq:Exp_CESH}
\end{align}
After the series expansion, we use the fact to obtain Eq.~\eqref{eq:Exp_CESH} that the integrand converges uniformly where one can swap the orders of the integration and the limit to infinity.
Note that 
\begin{equation}
\frac{2n + b - \widetilde{W}^{(\mathrm{C,H})}(x)^2 - (x-a_{\mathrm{L}})(a_{\mathrm{R}}-x)}{(x-a_{\mathrm{L}})(a_{\mathrm{R}}-x)} 
\end{equation}
equals zero if and only if $b=\beta=0$.
The radius of convergence for the expansion is thus
\begin{equation}
\left| \frac{2n + b - \widetilde{W}^{(\mathrm{C,H})}(x)^2 - (x-a_{\mathrm{L}})(a_{\mathrm{R}}-x)}{(x-a_{\mathrm{L}})(a_{\mathrm{R}}-x)} \right| = 1 ~.
\label{eq:ConvR_CESH}
\end{equation}

\begin{figure}
\centering
\includegraphics[scale=1]{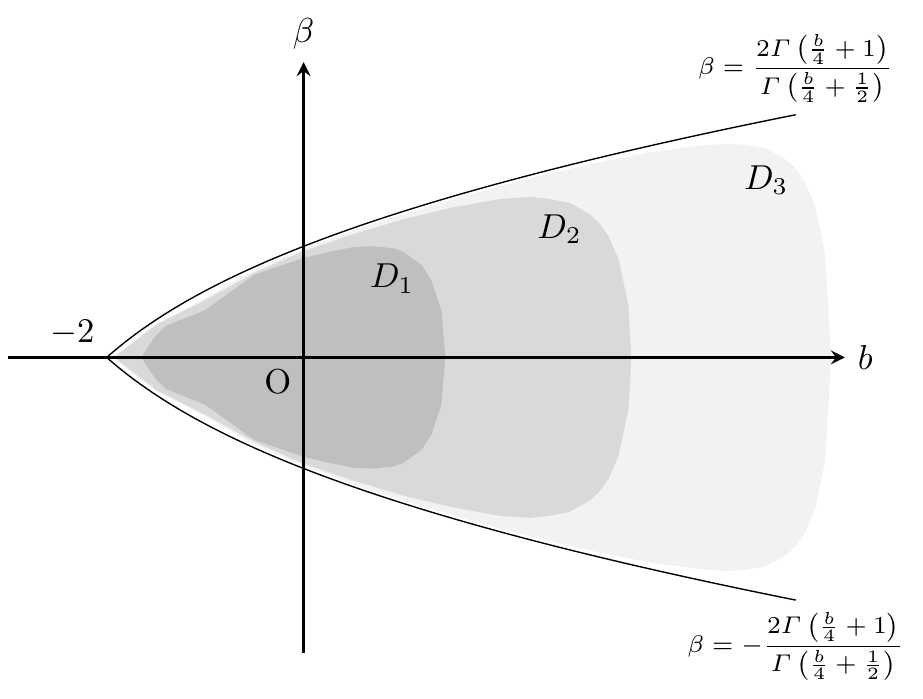}
\caption{The domains $D_n$ for $n=1,2,3$.
	Here, one can see that $D_1\subset D_2\subset D_3$.
	The parameters $b,\beta$ must satisfy Eqs.~\eqref{eq:cond1_CESH} and \eqref{eq:cond2_CESH} by construction.}
\label{fig:CESH_ExpDomain}
\end{figure}

Similarly, one can consider the expansion formulae for the cases of $\ast=\mathrm{L,J}$:
\begin{align}
I^{(\mathrm{C,L})} 
&\cong \frac{\pi}{4} \left( \sqrt{a'_{\mathrm{R}}} - \sqrt{a'_{\mathrm{L}}} \right) \nonumber \\
&\qquad\quad + \sum_{k=1}^{\infty}\frac{(-1)^k(2k)!}{(1-2k)(k!)^24^k} \int_{a'_{\mathrm{L}}}^{a'_{\mathrm{R}}} \frac{ \left\{ \left[ n+\frac{b}{4} - \widetilde{W}^{(\mathrm{C,L})}(z)^2 \right] z - (z-a'_{\mathrm{L}})(a'_{\mathrm{R}}-z) \right\}^k}{\big[ (z-a'_{\mathrm{L}})(a'_{\mathrm{R}}-z) \big]^{k-\frac{1}{2}}}\,\frac{\mathrm{d}z}{z} ~, 
\label{eq:Exp_CESL} \\
I^{(\mathrm{C,J})} 
&\cong \frac{2n+g+h}{2} \Bigg\{ \frac{\pi}{2} \left( 2 - \sqrt{(1-a'_{\mathrm{L}})(1-a'_{\mathrm{R}})} - \sqrt{(1+a'_{\mathrm{L}})(1+a'_{\mathrm{R}})} \right)  \nonumber \\
&~ + \sum_{k=1}^{\infty}\frac{(-1)^k(2k)!}{(1-2k)(k!)^24^k} \int_{a'_{\mathrm{L}}}^{a'_{\mathrm{R}}} \frac{ \left[ \frac{4n(n+g+h)+b - \widetilde{W}^{(\mathrm{C,J})}(y)^2}{(2n+g+h)^2} \left( 1-y^2\right) - (y-a'_{\mathrm{L}})(a'_{\mathrm{R}}-y) \right]^k}{\big[ (y-a'_{\mathrm{L}})(a'_{\mathrm{R}}-y) \big]^{k-\frac{1}{2}}}\,\frac{\mathrm{d}y}{1-y^2} \Bigg\} ~.
\label{eq:Exp_CESJ}
\end{align}
For the series \eqref{eq:Exp_CESL} and \eqref{eq:Exp_CESJ} to be convergent,
\begin{gather}
\left| \frac{ \left[ n+\frac{b}{4} - \widetilde{W}^{(\mathrm{C,L})}(z)^2 \right] z - (z-a'_{\mathrm{L}})(a'_{\mathrm{R}}-z)}{(z-a'_{\mathrm{L}})(a'_{\mathrm{R}}-z)} \right| \leqslant 1 ~, 
\label{eq:ConvR_CESL} \\
\left| \frac{ \frac{4n(n+g+h)+b - \widetilde{W}^{(\mathrm{C,J})}(y)^2}{(2n+g+h)^2} \left( 1-y^2\right) - (y-a'_{\mathrm{L}})(a'_{\mathrm{R}}-y)}{(y-a'_{\mathrm{L}})(a'_{\mathrm{R}}-y)} \right| \leqslant 1 ~,
\label{eq:ConvR_CESJ}
\end{gather}
respectively.

The radius of convergence for the series \eqref{eq:Exp_CESH}, \eqref{eq:Exp_CESL} and \eqref{eq:Exp_CESJ} are given by Eqs.~\eqref{eq:ConvR_CESH}, \eqref{eq:ConvR_CESL} and \eqref{eq:ConvR_CESJ}.
In terms of Eqs.~ \eqref{eq:Exp_CESH}, \eqref{eq:Exp_CESL} and \eqref{eq:Exp_CESJ}, a choice of parameters $(b,\beta)$ within this radius of convergence corresponds to a CES system which  is connected to the original conventional SI potential.
When a choice of parameters $(b,\beta)$ is outside the radius, such system simply does not relate to the original conventional SI potential in terms of those series.
We plot the domains where the series converges for $n=1,2,3$ on $(b,\beta)$-plane in Fig.~\ref{fig:CESH_ExpDomain}.
One can obtain the domains by solving Eq.~\eqref{eq:ConvR_CESH} numerically.
Let us call each domain $D_n$ respectively.
We have checked numerically that as $n$ grows, the radius of convergence is enlarged; $D_1 \subset D_2 \subset D_3 \subset\cdots$.
A quantitative argument of the inclusion relation of domains $D_n$ supports this result.
Thus we conclude that there always exist sets of model parameters where the expansion \eqref{eq:Exp_CESH} is possible for any $n$, and it is enough to consider the domain $D_1$ so that the expansion formula \eqref{eq:Exp_CESH} holds for any $n$.

\subsection{Numerical analysis on the residual}

\begin{figure}
\centering
\hspace*{-.05\linewidth}
	\begin{minipage}[t]{0.45\linewidth}
	\centering
        \subfloat[][$\beta=0$.]{
          \includegraphics[scale=0.63]{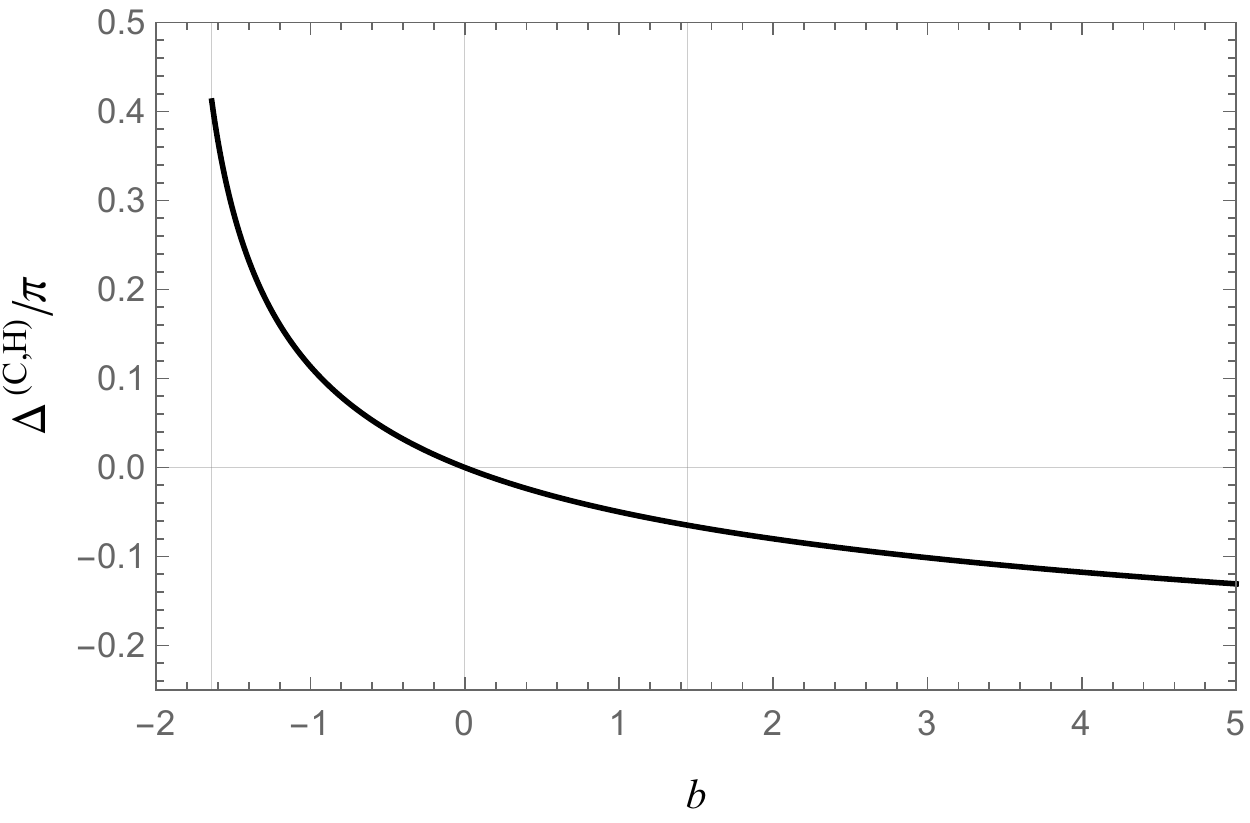}
          \label{fig:Delta_CESH_numB_n1}
        }
	\end{minipage}
\qquad\quad
	\begin{minipage}[t]{0.45\linewidth}
	\centering
        \subfloat[][$b=0$.]{
          \includegraphics[scale=0.63]{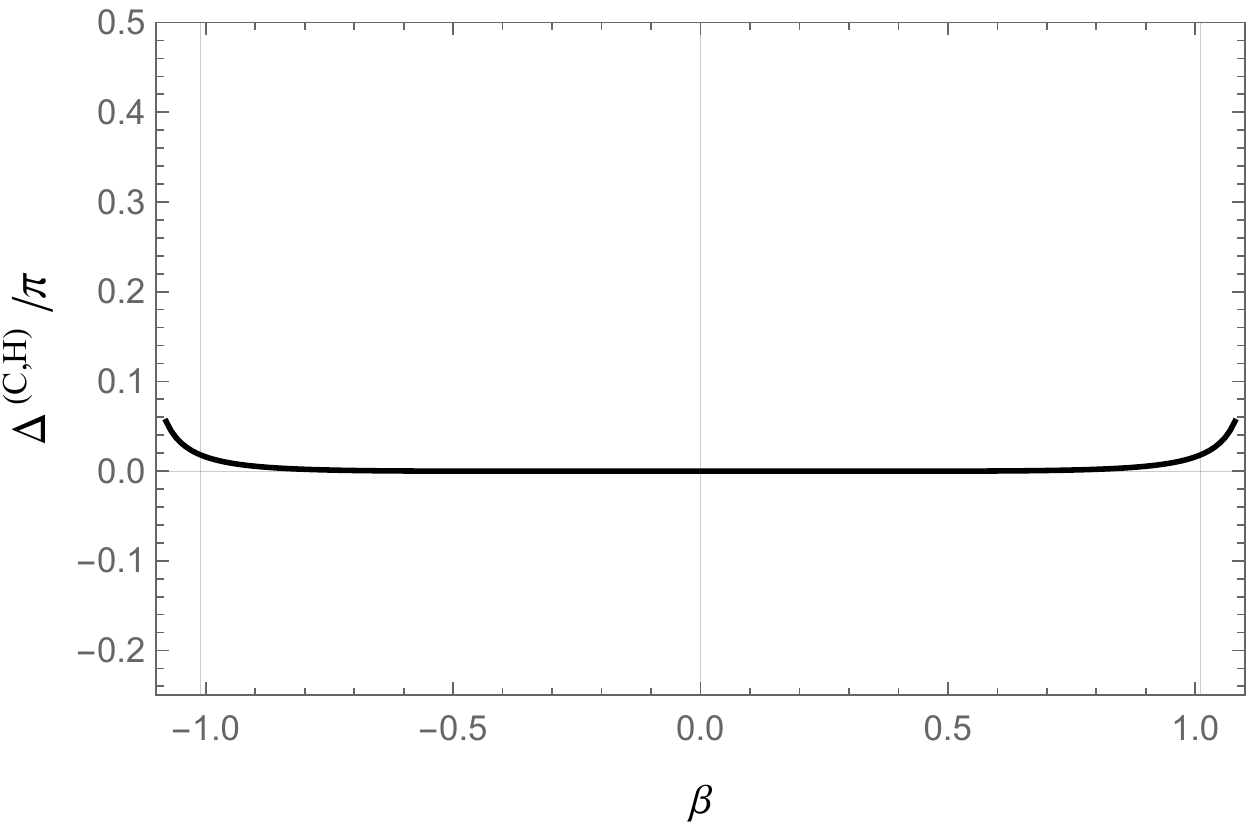}
          \label{fig:Delta_CESH_numBeta_n1}
        }
        \end{minipage}
\caption{Plots of $\Delta^{(\mathrm{C,H})}$ for $n=1$ (numerical results).
	The range of each plot is determined so that the systems will not have more than two turning points.
	The parametric conditions \eqref{eq:cond1_CESH} and \eqref{eq:cond2_CESH} yield $b>-2$ and $|\beta| < 2/\sqrt{\pi}$ respectively.
	The light gray lines at (a) $b\approx -1.64, 1.44$, (b) $|\beta|\approx 1.01$ show the radius of convergence \eqref{eq:ConvR_CESH}.}
\label{fig:Delta_CESH_num}
\end{figure}

We evaluate the residual 
\begin{equation}
\Delta \coloneqq n\pi\hbar - I_{\mathrm{SWKB}} ~,~~~
\Delta^{(\mathrm{C,\ast})} \coloneqq n\pi - I^{(\mathrm{C},\ast)}
\end{equation} 
as a function of $b$ and $\beta$.
Since our initial aim is to investigate the correction terms of the SWKB condition perturbatively, it would be relevant when we express the residual $\Delta$ in a series.
By employing our formulation, which is basically $I_{\mathrm{SWKB}}\cong\sum_{k=0}^{\infty}I^{(k)}$, $\Delta$ can be evaluated in the form of a series: $\Delta \cong n\pi\hbar - \sum_{k=0}^{\infty}I^{(k)}$.
Especially for $\ast=\mathrm{H}$, by using Eq.~\eqref{eq:Exp_CESH}, 
\begin{align}
\Delta^{(\mathrm{C,H})}
&\cong \left[ n - \frac{(a_{\mathrm{R}} - a_{\mathrm{L}})^2}{8} \right] \pi \nonumber \\
&\hspace{.075\textwidth} - \sum_{k=1}^{\infty}\frac{(-1)^k(2k)!}{(1-2k)(k!)^24^k} \int_{a_{\mathrm{L}}}^{a_{\mathrm{R}}} \frac{ \left[ 2n + b - \widetilde{W}^{(\mathrm{C,H})}(x)^2 - (x-a_{\mathrm{L}})(a_{\mathrm{R}}-x) \right]^k}{\left[ (x-a_{\mathrm{L}})(a_{\mathrm{R}}-x) \right]^{k-\frac{1}{2}}}\,\mathrm{d}x ~.
\label{eq:Delta_CESH}
\end{align}
All terms in Eq.~\eqref{eq:Delta_CESH} vanish when $b=\beta=0$, while no term is equals to zero for other cases.
We display the numerical calculation of $\Delta^{(\mathrm{C,H})}$ in Fig.~\ref{fig:Delta_CESH_num}.
First, we fix $\beta=0$ and see $\Delta^{(\mathrm{C,H})}$ as a function of $b$ (Fig.~\ref{fig:Delta_CESH_numB_n1}).
The value of $\Delta^{(\mathrm{C,H})}$ declines as $b$ grows.
$\Delta^{(\mathrm{C,H})}$ is negative when $b>0$, while $\Delta^{(\mathrm{C,H})}$ is positive for $b<0$.
Note that for the case of $b=0$, \textit{i.e.}, 1-d H.O., $\Delta^{(\mathrm{C,H})}$ equals zero.
Next, we fix $b=0$ and see $\Delta^{(\mathrm{C,H})}$ as a function of $\beta$ (Fig.~\ref{fig:Delta_CESH_numBeta_n1}).
$\Delta^{(\mathrm{C,H})}$ shows plateau behavior around $\beta=0$, and it grows around $|\beta|\approx 2/\sqrt{\pi}$.

\begin{figure}
\centering
\hspace*{-.05\linewidth}
	\begin{minipage}[t]{0.45\linewidth}
	\centering
        \subfloat[][$\beta=0$.]{
          \includegraphics[scale=0.63]{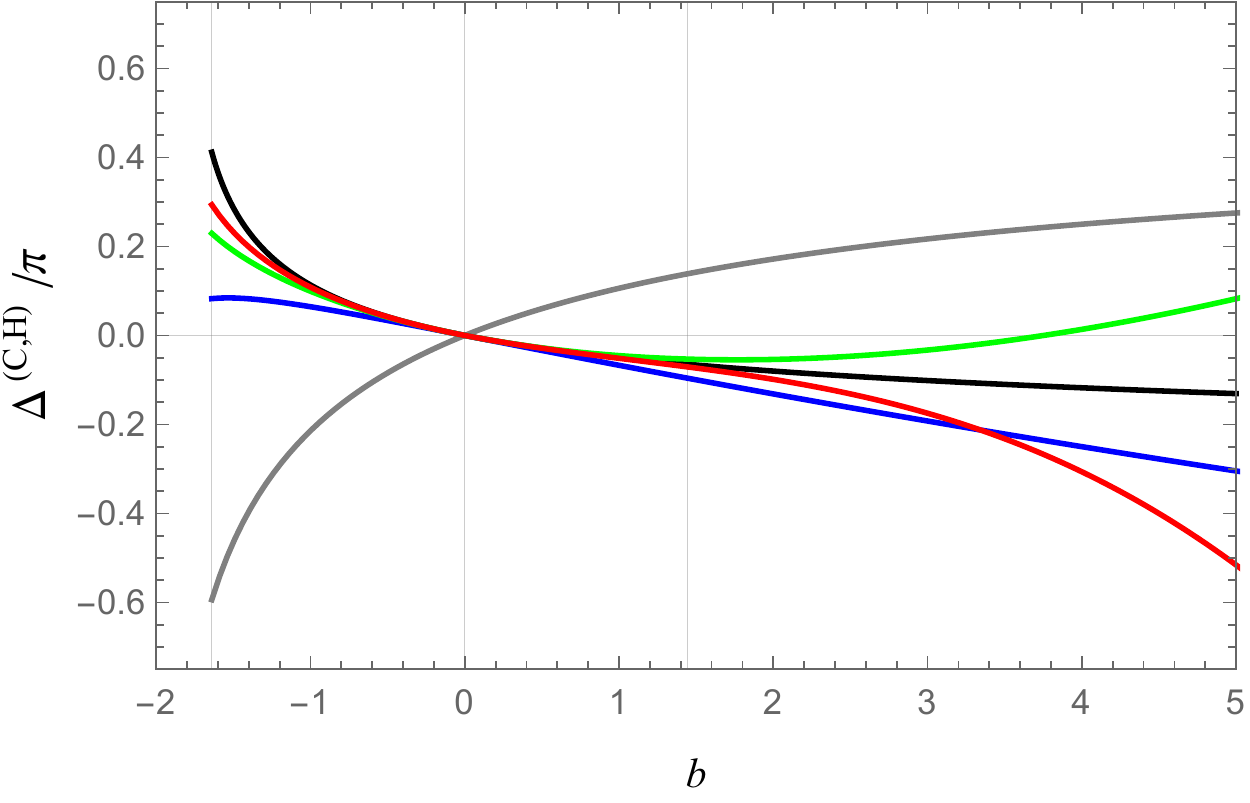}
          \label{fig:Delta_CESH_Exp_beta0b_n1}
        }
	\end{minipage}
\qquad\quad
	\begin{minipage}[t]{0.45\linewidth}
	\centering
        \subfloat[][$b=0$.]{
          \includegraphics[scale=0.63]{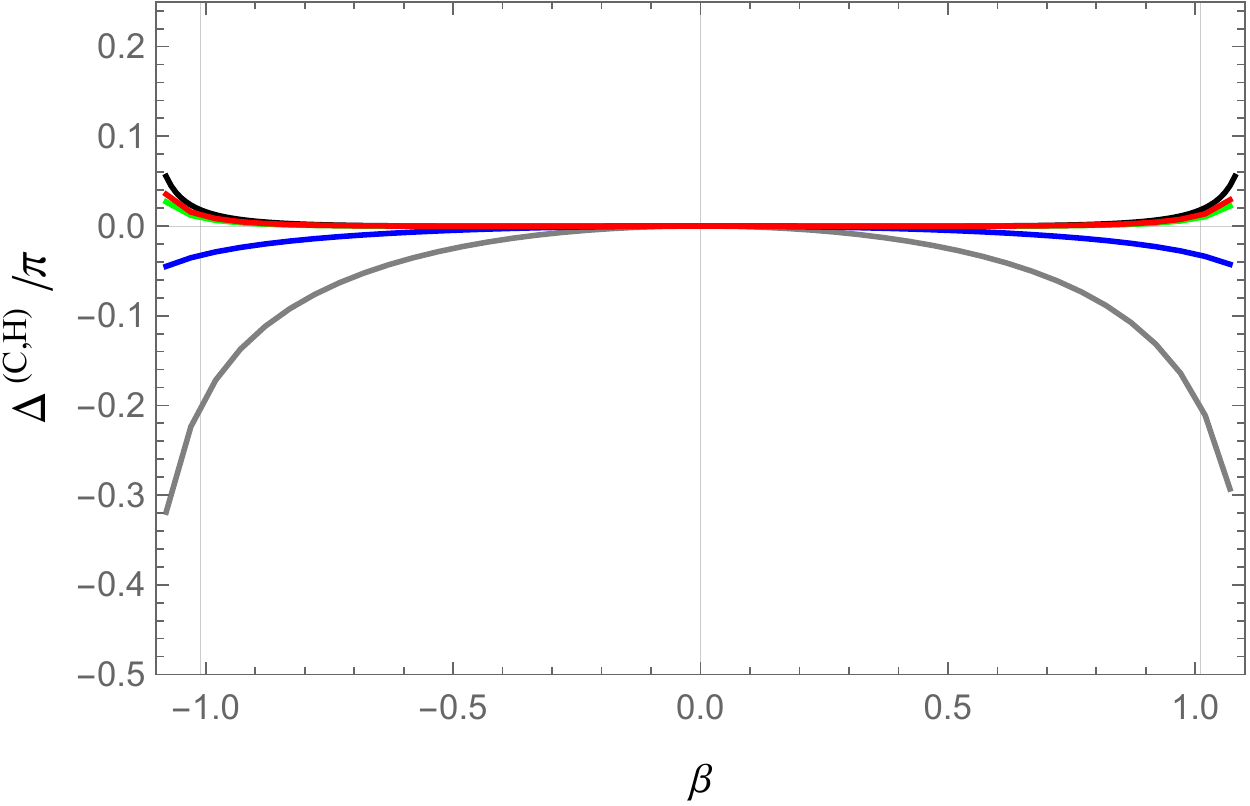}
          \label{fig:Delta_CESH_Exp_b0beta_n1}
        }
        \end{minipage}
\caption{Plots of $\Delta^{(\mathrm{C,H})}$ with different orders of the power series approximation for $n=1$. 
	We plot the 0th order of the r.h.s. of Eq.~\eqref{eq:Exp_CESH} (gray), up to first order (blue), second order (green) and third order (red), while the black curve shows the numerical result $\Delta^{(\mathrm{C,H})}$.
	The range of each plot is determined so that the systems will not have more than two turning points.
	The parametric conditions \eqref{eq:cond1_CESH} and \eqref{eq:cond2_CESH} yield $b>-2$ and $|\beta| < 2/\sqrt{\pi}$ respectively.
	The light gray lines at (a) $b\approx -1.64, 1.44$, (b) $|\beta|\approx 1.01$ show the radius of convergence \eqref{eq:ConvR_CESH}.}
\label{fig:Delta_CESH_Exp}
\end{figure}

In Fig.~\ref{fig:Delta_CESH_Exp}, we plot $\Delta^{(\mathrm{C,H})}$ with different orders of the power series approximation.
For glancing behavior of $\Delta^{(\mathrm{C,H})}$, the first few orders of the expansion formula are sufficient. 
Similar analysis can be done for the cases of $\ast = \mathrm{L,J}$.

\section{Conclusion}
\label{sec:concl}

In this paper, we have studied the non-exactness of the SWKB condition for the CES systems by Junker and Roy.
First we compared the singularity structures of the SWKB integrand and the QMF.
They are different, and thus one can deduce that the SWKB condition does not reproduce the exact bound-state spectra for this class of potentials.
For the CES systems, we found that the singularity structures possess the following properties.
The QMF has $n$ poles on the real axis and infinite number of poles on the complex plane other than that, whose effect on the contour integral vanishes pairwisely.
This feature proves the quantization condition in the QHJ formalism to be exact.
On the other hand, the SWKB integrand does not have the pairwise-cancellation property.
It has infinite number of poles on the complex plane, which are not to be treated analytically.
More than that, branch cuts spread all over the complex plane, which is also impossible to evaluate the effect on the contour integral.
One can see the non-exactness of the SWKB condition for the CES systems comes from the above properties.
Then, we numerically confirmed that the condition equation is not an exact one for the systems.
Our analysis on the $b$- and $\beta$-dependency of the value of SWKB integral reveals that, as pointed out in our previous letter, the deviations of the SWKB condition relate to the modifications of the whole distribution of the energy eigenvalues.

Also, we have shed light on the residual $\Delta$ and introduced a novel way of evaluating it for a case of the CES system.
We employed a perturbative approach, where we chose the non-perturbed system as a conventional SI potential and expanded the SWKB integral in powers of a parameter.
Our formulation realizes the change of the value of SWKB integral, \textit{i.e.}, the level structure of a system, according to the change of model parameters describing the modifications of the level structure.
One can understand the behavior of the SWKB integral or the residual by a few simple integrations along with the real line.
Our approach unintentionally classifies the CES systems into two according to whether a CES system is inside or outside the radius of convergence for the expansion, \textit{i.e.}, whether or not a CES system connected to a conventional SI potential in terms of the series.
The physical interpretation of the radius of convergence is still an open question.

As was mentioned in Sec.~\ref{sec:Intro}, an ``exact'' SWKB formula, whose leading term corresponds to the current SWKB formalism surely exists.
We do not know what the exact formula may look like so far.
However, by virtue of the expansion formula, hopefully the ``unknown parameter'' will be identified, as a result of which ``exact'' SWKB is to be formulated.
We also mentioned in Sec.~\ref{sec:Intro} that the relation between the SWKB condition and the solvability of the Schr\"{o}dinger equation has been discussed.
We believe that our analysis on the residual $\Delta$ may give a clue to understand the inherent meaning of the SWKB condition in connection with the solvability of the Schr\"{o}dinger equation.

\appendix
\section{Energy spectrum from the SWKB quantization condition}
\label{sec:Appx}

One can deduce energy spectra from the condition equation analytically for the conventional SI potentials~\cite{hruska1997accuracy}. 
For other classes of potentials, one needs to compute numerically to get approximate energy spectra.
Here we demonstrate the procedure for the simplest case of $\ast = \mathrm{H}$ in Eq.~\eqref{eq:cSI_pot}. 
The superpotential is 
\[
W(x) = -\hbar\frac{\mathrm{d}}{\mathrm{d}x}\ln\left|\phi_0^{(\mathrm{H})}(x)\right| = \omega x
\]
in this case, and the SWKB integral is calculated as
\begin{equation}
I_{\mathrm{SWKB}} = \int_{-\sqrt{\mathcal{E}}/\omega}^{\sqrt{\mathcal{E}}/\omega} \sqrt{\mathcal{E} - \omega^2x^2} \,\mathrm{d}x 
= \frac{2\mathcal{E}}{\omega} \int_0^1 \sqrt{1-t^2} \,\mathrm{d}t
= \frac{\pi\mathcal{E}}{2\omega} ~,
\end{equation}
where $t\equiv \omega x/\sqrt{\mathcal{E}}$.
From  the SWKB quantization condition, one obtains 
\begin{equation}
I_{\mathrm{SWKB}} = \frac{\pi\mathcal{E}}{2\omega} = n\pi\hbar
\qquad\text{\it i.e.}\qquad
\mathcal{E} = 2n\hbar\omega ~,
\end{equation}
which indeed agrees with Eq.~\eqref{eq:cSI_Ene}.

\bigskip
\section*{Acknowledgment}

The authors would like to thank Ryu Sasaki for his careful and useful advice and comments.  
We also appreciate Naruhiko Aizawa and Atsushi Nakamula for valuable discussions. 
Discussions during the YITP workshop YITP-W-20-03 on ``Strings and Fields 2020'', and YITP workshop YITP-W-21-04 on ``Strings and Fields 2021'' have been useful to complete this work. 
Y.N. is supported by the Sasakawa Scientific Research Grant from the Japan Science Society (No. 2022-2011), and JST SPRING Grant Number JPMJSP2151.
N.S. is supported in part by JSPS KAKENHI Grant Number JP B20K03278.

\bibliography{SWKB_CES_Rev_bib.bib}

\end{document}